\documentclass{aa}
\makeatletter
\AtBeginDocument{%
  \let\linenumbers\relax
  \let\endlinenumbers\relax
}
\makeatother
\usepackage{graphicx}
\usepackage{lipsum}
\usepackage{lscape} 
\usepackage{placeins} 
\usepackage{xurl}
\usepackage{subcaption}
\usepackage[colorlinks=true,linkcolor=blue,citecolor=blue,urlcolor=blue]{hyperref}
\usepackage{footnotehyper}
\usepackage{tablefootnote}
\usepackage{longtable}
\usepackage[varg]{txfonts}
\usepackage{tabularx}
\usepackage{booktabs}
\usepackage{siunitx}
\usepackage{caption}
\usepackage{txfonts}
\usepackage{threeparttable}
\usepackage{hyperref}
\usepackage[normalem]{ulem}
\usepackage[dvipsnames]{xcolor}

\newcommand{\customcitep}[3]{(see \citeauthor{#1} \citeyear{#1} and \citeauthor{#2} \citeyear{#2} for \emph{Kepler} and \citeauthor{#3} \citeyear{#3} for TESS)}

\newcommand{\PLATORedBook}{\hyperlink{cite.PLATORed-Book}{PLATO Red Book (2017)}}

\newcommand{\corot}{CoRoT\xspace}
\newcommand{\kepler}{\emph{Kepler}\xspace}
\newcommand{\ktwo}{K2\xspace}
\newcommand{\tess}{TESS\xspace}

\newcommand{\plato}{PLATO\xspace}

\newcommand{\ntr}{{\rm n_{\rm tr}}}
\newcommand{\teb}{{\rm t_{\rm EB}}}
\newcommand{\onesigma}{$1$-\rm{$\sigma$}\xspace} 

\newcommand{\sprk}{{\rm SPR_{\rm k}}}                  
\newcommand{\sprkext}{{\rm SPR_{\rm k}^{\rm ext}}}  
\newcommand{\sprkmaxsec}{{\rm SPR_{\rm kmax}^{\rm sec}}}
     
\newcommand{\sprtot}{{\rm SPR_{\rm tot}}}               
\newcommand{\sprtotext}{{\rm SPR_{\rm tot}^{\rm ext}}}
\newcommand{\nsr}{{\rm NSR_{\rm 1h}}}               
\newcommand{\nsrsec}{{\rm NSR_{\rm 1h}^{\rm sec}}} 
\newcommand{\nsrext}{\mathrm{NSR}_{\mathrm{1h}}^{\mathrm{ext}}}

\newcommand{\deltap}{\rm \delta_{\rm p}}  
\newcommand{\deltaEB}{\rm \delta_{\rm EB}}
  
\newcommand{\etamin}{\rm \eta_{\rm min}}           
\newcommand{\etanom}{\rm \eta_{\rm k}^{\rm nom}}   
\newcommand{\etaext}{\rm \eta_{\rm k}^{\rm ext}}

\urldef{\myurl}\url{https://vizier.cds.unistra.fr/viz-bin/VizieR-4?-ref=VIZ669592242c1cff&-out.add=.&-source=I/355/gaiadr3&-c=097.20189127476%20-45.87593420116,eq=ICRS,rs=2&-out.orig=o&-out.all=1}

\begin{document}

    \titlerunning{FP detection methods for PLATO}

    \authorrunning{Guti\'errez-Canales et al.}

    \title{Detecting false positives with PLATO using double-aperture photometry and centroid shifts}

   \author{F. Guti\'errez-Canales
          \inst{1,2}
          ,
          R. Samadi
          \inst{1}
          , 
          A. Birch 
          \inst{2}
          ,
          J. Cabrera
          \inst{3}
          ,
          C. Damiani
          \inst{2}
          ,
          P. Guterman,
          \inst{4}
          C. Paproth
          \inst{3}
          ,
          M. Pertenais
          \inst{3}
          and
          A. Santerne
          \inst{4,5}
          }

\institute{LIRA, Observatoire de Paris, Universit\'e PSL, CNRS, Sorbonne Universit\'e, Universit\'e Paris Diderot, Sorbonne Paris Cit\'e, 5 place Jules Janssen, 92195, Meudon, France\\
              \email{fernando.canales@obspm.fr}
         \and Max Planck Institute for Solar System Research, Justus-von-Liebig-Weg 3, 37077 Gottingen, Germany
         \and Deutsches Zentrum für Luft- und Raumfahrt, Rutherfordstr. 2, 12489 Berlin, Germany 
         \and Aix Marseille Univ, CNRS, CNES, LAM, Marseille, France
         \and Univ. Grenoble Alpes, CNRS, IPAG, 38000 Grenoble, France}

   \date{Received 4 December 2024 / Accepeted 12 December 2025}

   \abstract      
   {PLATO will discover exoplanets around Sun-like stars through transits and characterize their host stars through asteroseismology. Since photometry for most PLATO targets will be extracted on-board, an efficient strategy to detect false positives (FPs) --transit-like signals not caused by planets-- is needed. Centroid shifts are a standard FP diagnostic. However, only 5\% to 20\,\% of PLATO's largest stellar sample (P5 sample) will have centroids computed on-board. An alternative on-board strategy is necessary for the remaining targets.}
   {We propose a double-aperture photometry strategy to detect FPs. We tested two mask types: extended masks, enlarging the nominal aperture, and secondary masks, centered on the main contaminant. For each, we derived flux and centroid-shift metrics to assess which best discriminates FPs.}
    {Using Gaia DR3, we defined our P5 targets and their background stars, which we assumed were eclipsing binaries (EBs) with transit depths and durations drawn from observed distributions. From simulated photometry and centroid shifts we computed extended and secondary fluxes; extended, secondary, and nominal centroids, then compared their FP detection efficiency.}
   {Under these assumptions, $\sim35\,$\% of P5 targets have a single FP-creating contaminant and $\sim22\,$\% have two or more. Extended centroid shifts reach 87~$\%$ efficiency, nominal 84~$\%$ and secondary 75~$\%$. Secondary flux attains 92~$\%$ efficiency while extended 73~$\%$. }
   {Secondary flux is the most efficient metric followed by extended, nominal, and secondary centroids, then extended flux. Since double-aperture photometry is 50$\%$ cheaper in CPU and telemetry budgets, secondary and extended fluxes are optimal for most P5 targets. Secondary masks are the best option for the P5 targets with a single FP-creating contaminant, while extended masks suit for P5 targets where extended flux could be as competitive as centroids. Our work shows that double-aperture photometry and centroid shifts will allow PLATO to correctly discard a very large fraction of FPs from EBs.}


\maketitle
    
\section{Introduction}
\label{sec:Intro}

The launch of the Hubble Space Telescope (HST) in 1990 revolutionized astronomy. Devoted to different scientific objectives regarding exoplanets, several space missions have followed the HST path. Among these are \corot \citep{2009Auvergne}, \kepler \citep{Borucki2007, Borucki2010} and \tess \citep{2015JATIS...1a4003R}. These missions enabled the discovery of almost 6000 exoplanets (and 7000 candidates) so far\footnote{As of 29th of July of 2025 from the \href{https://exoplanetarchive.ipac.caltech.edu/}{NASA Exoplanet Archive}}. In late 2026 the PLAnetary Transits and Oscillations of stars (\plato) mission \citep{2014ExA....38..249R,Rauer_Heras_2018,2025ExA....59...26R} will lead to major progress in this domain. With \plato , the number of Earth-like planets detection will increase. Furthermore, the characterization of such type of planets in terms of radius, mass and age will also be improved \citep[see][]{2022Rene}. 

In order to achieve  its objectives, \plato will use several stellar samples \citep[see][]{2021A&A...653A..98M}. Sample 1 (P1) is the backbone sample of the mission. It contains stars bright enough (lower than V=11, where the dominant noise source is the photon noise, with a maximum random noise, or random Noise-to-signal ratio, NSR, of 50 ppm in one hour) to allow ground-based radial velocity follow up and detailed characterization of the host stars thanks to asteroseismology. Sample 2 (P2) consists of stars brighter than V = 8.5 with the same spectral types and noise performance as the P1 stars. Sample 4 (P4) consists of nearby cool late-type dwarf stars with habitable zones relatively close-in, making the planets in their habitable zone to have orbital periods of only few weeks.
Sample 5 (P5), the largest one, is derived from the requirement of observing a large number of stars to obtain statistical information on planetary properties. Hence, with this sample more than 4,000 planets are expected to be detected \citep[see][]{2025ExA....59...26R}. However, only a small fraction of these stars can be characterized with asteroseismology \citep[see][]{2024A&A...683A..78G} and the mass of their planet determined with radial velocity (RV) observations. \plato has chosen to extract on-board photometric measurements of a large number of stars. For the brightest targets, imagettes (regions of few pixel side, postage-stamps in \kepler jargon) will be downloaded to ground unprocessed. However, for most of the targets, aperture photometry will be computed on-board (similarly to \corot, see \citealt{2019Marchioripaper}).

The \plato mission will generate over 100 Terabits of raw data daily, far exceeding the available bandwidth for data download \citep[see][]{2025ExA....59...26R}, that is of 435 Gbits/day. This huge data volume creates significant constraints on \plato’s on-board CPU and telemetry systems (for more details about the on-board data processing for \plato we refer to the work by \cite{10396312}).  As a result, the mission’s on-board software must prioritize essential analyses and computations. Given these constraints, the mission strategy will rely on collecting and processing data on-board for posterior ground-based validation, distinguishing transit signals that are unlikely to be false positives, FPs. By ``validation'' we mean assessing a transit signal against one or several criteria to see how likely it is for the signal to be planetary~\citep[e.g.][]{2011ApJ...727...24T,2014MNRAS.441..983D}. For instance, a common procedure is to  match the observed light curve to a planetary model rather than an Eclipsing Binary (EB) \citep[see][]{2012ApJ...750..112L}. The word ``validation'' should not be confused within this context with the word ``confirmation''. The confirmation of a transit signal as planetary is achieved typically by measuring the mass of the planet creating the signal, usually via RV measurements. Validation is particularly crucial for \plato, given that confirmation through RV observations is not possible given the huge number of target stars for the mission. For instance there will be up to $\sim 245,000$ P5 targets. Furthermore, only $4500$ of the brightest P5 targets will have their photometry extracted on-ground from imagettes. All these considerations show that \plato faces considerable constraints in on-board processing and telemetry due to the vast volume of data generated. Furthermore, even if it would be useful to know, it is not in the scope of this paper to realistically determine expected EB populations for \plato.

A well-known method for detecting as much as possible FPs, and therefore making possible to validate transit signals, is measuring centroid shifts. Centroid shifts are widely used to detect FPs in ground-based transit surveys \citep[see][]{2017Gunther} as well as on space-based transit surveys \customcitep{2010ApJ...713L.103B}{2013PASP..125..889B}{2021RNAAS...5..262H}\footnote{The introduction present in \citet{2024AJMelton} shows an extensive list of papers regarding the vetting of exoplanet candidates from transit signals.}. For \plato, centroid shift measurements were part of the initial FP detection strategy.  It was thus proposed to obtain centroids on-board for at least $5\,\%$ of the targets \citep{ESA_2021}. Due to technical limitations, no more than  $20\,\%$ of the targets  can have centroids measured on-board. Furthermore, since for most of the targets the photometry is extracted on-board, only the information of a very small fraction of targets is going to be downloaded to detect FPs on-ground with a series of techniques. Due to all these reasons, the concept of double-aperture photometry was proposed, since a strategy was needed for the targets that wouldn't have centroid shift measurements.

The current strategy to detect FPs for \plato involves a range of techniques, where two of them are the computation of centroid shifts and the use of the double-aperture photometry method. However, the efficiency of this new strategy has never been assessed until now. The purpose of this paper is to provide the first overall efficiency study regarding double-aperture photometry for \plato and its use alongside Centroid shifts, even if centroid shifts will be analyzed afterwards, on-ground. The idea of double-aperture photometry is to have two masks or apertures per CCD window focused on a given star. One of these masks is the nominal mask, which is centered in the target and it is used to extract the target photometry with the highest possible Signal-to-Noise-Ratio (SNR). The purpose of the additional mask is to measure an additional flux that, as will be shown here, enables to detect FPs under certain conditions. 

We consider two options for the additional mask. One option is the so-called extended mask and the other is the so-called secondary mask. The extended mask is an extension/expansion of the nominal mask, making it larger, in order to reach contaminant stars surrounding each target.  In change, the secondary mask is a smaller mask centered only on the most prominent contaminant around each target. With the extended mask we can obtain what we refer as to the extended flux (or extended photometry) and with the secondary mask we obtain the secondary flux (or secondary photometry). We can also use extended and secondary masks to compute centroid measurements. These centroids are alternatives to the centroids measured with the nominal mask. Similar approaches have appeared in the literature, for example in \cite{2017A&A...606A..75C} who applied comparable vetting techniques using \ktwo data. However, in the context of \plato, double-aperture photometry is advantageous as it requires 50$\,\%$ less CPU and telemetry than centroid shifts\footnote{As it requires only one measurement per imagette for flux and two for centroid computations (\rm x and \rm y positions) per imagette.}, making it an attractive alternative for on-board implementation. The approach followed in the present  work can be applied to future exoplanet missions based on transit detection. This approach allows to discard FPs in order to enhance the selection of true positives. By true positives we mean transit signals coming from real planets around target stars.

First, we give a small over-view of the \plato mission and photometry extraction methods in Section \ref{sec:PLATO_mission}. We present the proposed method of double-aperture photometry for detecting FPs and how to use it to perform flux measurements in Section \ref{sec:Double_Aperture}. The centroid method for detecting FPs is described in Sect. \ref{sec:Centroid}. Our analysis with all the corresponding assumptions is presented in Sect. \ref{sec:methods} and we show how we computed the efficiency of each method in Section \ref{sec:Efficiency}. We show the results in Sect. \ref{sec:Results} and our conclusions in Section \ref{sec:Conclusions}.

\section{PLATO mission}
\label{sec:PLATO_mission}
\subsection{PLATO instrument}
\label{sec:PLATO_inst}

\plato payload is composed of 26 cameras mounted on a single optical bench. Each camera is a  12~cm pupil diameter, wide-field refractive telescope that see $\sim $ 1037 $\rm deg^{2}$ \citep[see][]{2021SPIE11852E..09P, 2021SPIE11852E..4YP}. Out of the 26 cameras, 24 work at a cadence of 25~s and are called normal cameras (N-CAM). N-CAMs are used for the core science observations and are divided in four groups of six. The remaining 2 cameras work at a cadence of 2.5~s and are called fast cameras (F-CAM).  There are four Charge Couple Devices (CCDs) mounted on the focal plane of each N-CAM and F-CAM. Also, all N-CAMs of a given group  share the same Line-Of-Sight (LOS) and Field-Of-View (FoV). Every $\sim 91 $ days (this period of time is known as a ``quarter'') the spacecraft has to rotate 90 degrees to have the solar panels directed towards the sun \citep[see][]{Rauer_Heras_2018}. 

The expected time duration for \plato mission is four years, having the idea of two long observation phases (LOPS) of two years each or even one LOP of three years. If an extended mission is approved, a step-and-stare phase could be scheduled afterwards. As mentioned by \citet{2022A&A...658A..31N}, the center of both PLATO LOPS fields are inside the spherical caps of ecliptic coordinate $| \beta |  > 63^{\circ}$ and the size of each field is $\sim 2232$ square degrees. 

In order to fulfill the mission science objectives four stellar samples have been defined, as mentioned in Sect. \ref{sec:Intro}. For this work we only consider the P5 sample which is often called the ``statistical sample" since it will be used for planet frequency studies. This is also the sample studied by \citet{2019Marchioripaper} and \citet{2023MNRAS.518.3637B}. The P5 sample contains at least 245,000 dwarf and sub-giant stars (F5-K7) with V $\leq$ 13 mag and temperatures ranging from 3875~\rm{K} to 6775~\rm{K} \citep{2023MNRAS.518.3637B}. The number of targets of this samples assumes two long duration observation phases.
\subsection{PLATO photometry}
\label{sec:Plato_photo}

All PLATO photometric measurements, either done on-board or on-ground, rely on the concept of imagette. An imagette is a CCD window that surrounds a target star. The center of the imagette is located at no more than 0.5 pixels from the star barycenter \citep{march2019}. In the case of the P5 sample the imagettes will be 6~$\times$~6 CCD pixel squares. In particular, for the brightest targets the imagettes will be downloaded to extract their photometry on-ground. In this work we focus on the remaining P5 targets where the photometry will be extracted on-board, which are the vast majority of P5 targets and that is about 80,000 stars per camera. For each one of these targets a stellar light curve will be produced on-board the satellite at a cadence of 600~s or 50 s using an aperture mask.  It is important to mention now that there is a formal distinction between the words window and imagette for \plato pipeline. Briefly, the word imagette refers to the 6 by 6 pixel squares that are downloaded in order to extract their photometry and centroids on-ground using a PSF fitting method. While the word window refers to the 6 by 6 squares where we compute the flux and centroid on-board using an aperture.

\subsection{Nominal On-board photometry}
\label{subsec:nominal_on-board_photometry}
On-board photometry extraction is done by integrating the flux over a subset of the window pixels called the aperture or the mask. \citet{2019Marchioripaper}, in sect. 4.6.3, showed that binary masks are the best option for \plato and gave a procedure for obtaining a binary mask for every target. The binary mask obtained following this procedure is called the nominal mask. Furthermore, any time we refer to on-board photometry we are implying on-board photometry for N-CAM. 

The main concept behind the \citet{2019Marchioripaper} procedure is the noise-to-signal ratio, NSR (the inverse of the signal-to-noise ratio, SNR), of individual window pixels. The procedure of \citet{2019Marchioripaper} is summarized in Appendix~\ref{sec:Apendix_B}.  Fig.~\ref{fig:nominal} shows a schematic view of an example nominal mask, $\omega_{\rm n}$, obtained following this procedure. The NSR for a single \plato pixel is
\begin{equation}
    \rm NSR = \frac{\rm Noise}{\rm Signal} \; ,
    \label{eqn:NSR_simple}
\end{equation}
where by Signal we mean the flux of the target and by Noise we mean the overall noise in a given window. For PLATO light curves computed using a binary mask $\omega$ and in our case a single camera, the NSR is
\begin{equation}
    \rm NSR_{*} = \frac{ \sqrt{ \sum\limits_{\rm n=1}^{36} \left(  {\rm I^{T}_{n} } + \sum\limits_{\rm k=1}^{\rm N_{C}} { \rm I^{k}_{n}}  +  {\rm B} \, \Delta \rm t_{\rm exp} + \sigma^{2}_{\rm D} + \sigma^{2}_{\rm Q} \right) \omega_{n}}} {\sum\limits_{n=1}^{36} I_{n}^{T} \omega_{n}} \;.
    \label{eqn:NSR_11}
\end{equation}
where the subscript $\rm k$ runs over the number of contaminants of the imagette, $\rm N_{c}$, the subscript $\rm n$ refers to the imagette pixels $\{1, 2, ..., 36\}$ and $\rm \omega_{n}$ refers to the binary mask over those pixels. Eq.~(\ref{eqn:NSR_11}) gives the NSR value for each flux measurement involving an imagette based on the noise and signal contribution of each pixel. The most recent study about the noise budget for PLATO can be found in \cite{Borner2024}. Table \ref{tab:symbols} gives a complete description of the parameters in Eq. (\ref{eqn:NSR_11}).
For the noise related to the background flux (zodiacal light), we are using the median value from the distribution of background noise levels present in Fig. 8 of \cite{2019Marchioripaper}. We made this choice after using the lowest ($\rm 25 e^{-}/px/s$) and highest ($\rm 65 e^{-}/px/s$) values of the distribution and realizing only a small change in our results after using both values.

For a one-hour duration signal and a given number of \plato cameras, $\rm N_{T}$, we can use the following expression for the NSR in ppm $\cdot$ hr$^{1/2}$ (assuming NSR scales with multiple independent measurements)
\begin{equation}
    \rm NSR_{1h} = \frac{10^{6}}{12 \sqrt{\rm N_{T}}} NSR_{*}\;,
    \label{eqn:nsr_1h_nt_cameras}
\end{equation}
where the 12 constant refers to the square root of the number of samples in one hour for a given \plato N-CAM,i.e. using the 25~s cadence ($\rm \sqrt{3600~s/25~s} = 12$).

\begin{table}[htbp]
\captionsetup{singlelinecheck=false, justification=raggedright}
\caption{Noise and other parameters}
\label{tab:symbols}
    \begin{tabular}{|c | c | c |}
    \hline
    Symbol & Description & Value  \\
    \hline 
         $\rm I_n^{T}$ & Target star flux contribution & \\ 
                     &   to pixel intensity & \\
\hline
    $\rm I_n^{k}$      & Contaminant star flux &\\
                         & contribution to pixel & \\
                         & intensity & \\
    \hline
   $\rm B$   & Sky-background (zodiacal light) & $\rm  45 \mathrm{e^-}/px/s $\\
    \hline
    $\sigma_{\rm D}$ & Detector noise (readout, & \\
                         & smearing and dark current) &  $\rm 50.2 \mathrm{e^-}/px$ \\
    \hline
    $\sigma_{\rm Q}$ & Quantization noise &  $7.2 \rm \mathrm{e^-}/px $\\
    \hline
    $ \Delta \rm  t_{\rm exp}$ & Exposure time & 21~s\\
        \hline
    \end{tabular}
    \tablefoot{Description of terms in Eqs. (\ref{eqn:NSR_simple}) and (\ref{eqn:NSR}) and assumed values taken from \citet{2019Marchioripaper}.}
\end{table}

We now briefly introduce the concept of PLATO data products, which represent the final outputs of the PLATO mission. These are organized into four levels (0–3). Level 0 includes imagettes, raw light curves, and raw centroids computed on-board, while Level 1 contains processed versions of these. Level 2 involves asteroseismic parameter measurements as well as the transit planetary candidates. Finally, Level 3 data products consist of the ``final catalog'' of confirmed planetary systems. For this paper we assumed that L1 pipeline has perfectly removed the systematics and background flux of each imagette. In particular, we assumed there is no residual drift of the stars across the CCD. By ``residual drift'' we mean the remaining positional drift after the L1 correction. While drift and correction effectiveness do vary across the FoV, detailed assessment of this variation is ongoing work within the consortium.

\subsection{Detectability of transit signals}
\label{subsec:detectability_of_transit_signals}
In every imagette several stars surrounding the target will be present. Following the \citet{2019Marchioripaper} recommendation we refer as to  ``contaminant" any foreground star located within a 10~pixel radius from the target. \citet{2019Marchioripaper}  showed that above a distance of 10~pixels the probability for a contaminant to generate FP is very low. Contaminants that are EBs might produce signals that can be misinterpreted as planetary transits on the target. These events are the most common examples of FPs for \plato. In order to know the flux contribution of each contaminant to the total flux of the window, \cite{2019Marchioripaper} introduced the Stellar-Pollution Ratio ($\rm SPR$).
\begin{equation}
  \rm SPR_{k} = \frac{F_{k}}{F_{tot}} \; ,
  \label{eqn:spr_k}
\end{equation}
where $\rm F_{k}$ is the the flux of a single contaminant of index \rm{k} and $\rm F_{\rm tot}$ is the total flux (from all sources) present in the window. These terms are
\begin{equation}
    \rm F_{k} = \sum\limits_{n=1}^{36}  I_{n}^{k}  \omega_{n} \; ,
    \label{eqn:F_k}
\end{equation}
\begin{equation}
    \rm F_{tot} = \sum\limits_{n=1}^{36} {I}_{n} \, \omega_{n} \; , \label{eqn:F_tot}
\end{equation}
where  $\omega_{n}$ is the nominal mask and where $\rm {I}_{n}$ is the sum of the target contribution and the contribution of all contaminant stars:
\begin{equation}
        \rm {I}_{n}   =   {I}_{n}^T +   \sum\limits_{\rm k=1}^{\rm N_{C}} { \rm {  I}^{k}_{n}} \; .
        \label{eqn:I_n}
\end{equation}
We assume that the sky background has been removed. However, the noise from the background flux removal remains in the expression for the NSR in Eq. (\ref{eqn:NSR_11}). 
The total amount of pollution (i.e. total flux contribution from contaminants) in a given window is the sum of all the individual $\rm SPR_{k}$. This metric is named the total Stellar-Pollution-Ratio $\rm SPR_{tot}$:
\begin{equation}
 \label{eqn:sprt_tot}
 \rm SPR_{tot} = \sum\limits_{k = 1}^{N_{C}} SPR_{k} \; .
\end{equation}
The main feature of any transit signal is its transit depth, $\rm \delta$. This is the fraction of the area of the stellar disc that is covered during the transit. In the case of a planet transiting a star, the transit depth is called $\rm \delta_{p}$ and can be approximated as
\begin{equation}
    \delta_{\rm p} = \left( \frac{R_{\rm p}}{R_{*}}\right)^{2} \;,
    \label{eqn:transit_depth_planet}
\end{equation}
where $\rm R_{\rm p}$ is the radius of the planet and $\rm R_{*}$ is the radius of the star. Since we are interested in detecting FPs, we focus on the transit depth of individual background EBs, $\rm \delta_{\rm EB} $, instead of focusing on $\rm \delta_{\rm p}$.  We recall now that a given transit depth can also be defined in terms of flux. In the case of a transit happening on a contaminant star, $\rm k$, the transit depth is:
\begin{equation}
    \deltaEB = \frac{\rm F_{\rm k} - \rm F_{\rm k}^{\rm in}}{\rm F_{\rm k}} = \frac{\Delta \rm F_{\rm k}}{\rm F_{\rm k}}\; ,
    \label{eqn:transit_depth_flux}
\end{equation}
where $\rm F_{k}$ is the flux of the star out of transit and $\rm F_{k}^{in}$ is the flux of the star during a transit $ \left( \rm i.e.\, \rm F_{k}^{in} = \sum\limits_{n=1}^{36} I_{n}^{k, in} \right)$. The flux out of transit in the window is given by Eq.~(\ref{eqn:F_tot}) and can be re-expressed as
\begin{equation}
    \rm F^{\rm out} = \rm F_{\rm tot} = \rm F_{T} + \sum\limits_{\rm k=1}^{\rm N_{\rm C}}{\rm F_{\rm k}}\;.
    \label{eqn:F_tot_out_of_transit}
\end{equation}
The flux in the window when a transit is happening in a contaminant of index $\rm k$ is
\begin{equation}
    \rm F^{\rm in} = \rm F_{\rm tot} +   \rm F_{\rm k}^{\rm in} - \rm F_{\rm k}\; .
    \label{eqn:F_in_transit_window}
\end{equation}
The difference between Eq.~(\ref{eqn:F_tot_out_of_transit}) and Eq.~(\ref{eqn:F_in_transit_window}) is called $\rm \Delta F^{nom}$ where the superscript ``nom'' refers to the nominal mask and is 
\begin{equation}
    \Delta \rm F^{\rm nom} = \rm \delta_{\rm EB} F_{\rm k} \;,
    \label{eqn:delta_flux_window}
\end{equation}
where $\rm \delta_{\rm EB}$ (Eq.~(\ref{eqn:transit_depth_flux})) is the intrinsic transit depth (in all the following we assume $\rm \delta_{\rm EB}$  to be given in ppm). The quantity $\rm \delta_{\rm EB}$ is also diluted by a given amount due to the flux of the target and other contaminants in the window. This causes that we are not able to measure $\rm \delta_{\rm EB}$ directly, but what we call the ``observed'' (or ``apparent'') transit depth. This quantity is denoted $\rm $ $\rm\delta_{\rm k}^{\rm nom}$ and is obtained by dividing Eq.~(\ref{eqn:delta_flux_window}) over Eq.~(\ref{eqn:F_tot_out_of_transit})
\begin{equation}
    \rm \delta_{k}^{nom} = \delta_{\rm EB} \,  SPR_{k} \;.
    \label{eqn:apprent_transit_depth}    
\end{equation}
With this we are able to define the statistical significance of the transit. First we recall that the statistical significance, $\rm \eta$, of the transit is the ratio of the apparent transit depth, $\rm \delta_{k}^{nom}$, over the noise. The expression for the statistical significance in the nominal mask for a transit happening in a contaminant star is

\begin{equation}
    \eta_{\rm k}^{\rm nom} = \frac{\rm  \delta_{k}^{nom} \, F_{tot} \sqrt{\teb \, \rm \ntr}} { \sigma_{\rm F_{\rm tot}} }    \;,
    \label{eqn:bt_eta_raw_0}
\end{equation} 
where $\rm \teb$ and $\rm \ntr$ are, respectively, the transit duration (in hours) and number of transits, and $\rm \sigma_{F_{tot}}$ is the \onesigma dispersion in the light-curve averaged over 1 hour and over $\rm N_T$ cameras. We do this 1-hour average because we are using transit durations expressed in hours and also because measurements have to be averaged over the duration of a transit, that typically last only a few hours. 
By definition we have (see Eq.~(\ref{eqn:NSR_simple})):
\begin{equation}
\rm NSR_{1 hr} \equiv \frac{\sigma_{F_{\mathrm{tot}}}}{F_{T}} \; .
\label{eqn:NSR_1h_definition}
\end{equation}
Furthermore 
\begin{equation}
 \frac{\rm F_{\rm T}}{\rm F_{\rm tot}}   =  1 - \rm SPR_{tot} \; , 
 \label{eqn:F_T_over_F_tot}
\end{equation}
Accordingly, with the  help of Eqs.~(\ref{eqn:apprent_transit_depth}), (\ref{eqn:NSR_1h_definition}) and (\ref{eqn:F_T_over_F_tot}), Eq.~(\ref{eqn:bt_eta_raw_0}) can be rewritten as
\begin{equation}
    \eta_{\rm k}^{\rm nom} = \frac{\rm \delta_{\rm EB} \,  SPR_{k} \, \sqrt{\rm \teb \, \rm \ntr}}{(1 - \rm SPR_{tot}) \, \rm NSR_{1h}}  \;,
    \label{eqn:bt_eta}
\end{equation} 
where $\rm NSR_{1h}$ is given by Eq.~(\ref{eqn:nsr_1h_nt_cameras}). As can be seen, the significance scales with $\rm \delta_{EB} \sqrt{ \rm t_{EB}}$, which value varies depending on the star.

In order to be detectable, any transit signal should have a statistical significance value above a given threshold. We call this threshold $\etamin$. Based on considerations given by \citet{2010ApJ...713L.120J}, \citet{2019Marchioripaper} used $\eta_{\min} = 7.1$ for their analysis. More recent studies~\citep[e.g.][]{2018AJ....155..205H,2020AJ....160..159C,2020AJ....159..248K,2020AJ....159..279B} have challenged this choice. However, we keep it for consistency with the study by \citet{2019Marchioripaper}.
It follows that a contaminant star that is an EB generates a significant transit in the nominal flux (FP) whenever  we have
\begin{equation}
    \eta_{\rm k}^{\rm nom} > \eta_{\rm min} \; .
    \label{eqn:falsepos_etas}
\end{equation}
Following  \cite{2019Marchioripaper}, it is possible to derive a threshold in terms of  Stellar-Pollution-Ratio (SPR)  above which a transit in a contaminant star can cause a FP.  Indeed, using Eq.~(\ref{eqn:bt_eta}) and Eq.~(\ref{eqn:falsepos_etas}) it can been shown that this happens when
\begin{equation}
    \rm SPR_{k} > \rm SPR_{k}^{crit} \; ,
    \label{eqn:falsepos_spr}
\end{equation}
where 
\begin{equation}
    \rm SPR_{k}^{crit} = \frac{\eta_{min} (1 - \rm SPR_{tot}) \rm NSR_{1h}}{\delta_{\rm EB}  \sqrt{\teb \ntr}} \; ,
\label{eqn:sprcrit}
\end{equation}
is by definition the critical Stellar-Pollution-Ratio.

For completeness, we write the statistical significance in the nominal mask of a planet transiting the target star
\begin{equation}
\eta = \frac{\deltap \sqrt{\rm t_{\rm p} \ntr}}{\rm NSR_{1h}} \; , 
\label{eqn:significance_true_planet}
\end{equation}
where $\rm \delta_{p}$, $\rm t_{p}$ and $\ntr$ are, respectively, the planet transit depth (in ppm), the transit duration in hours and the observed number of transits (metrics like Eq.(\ref{eqn:significance_true_planet}) are called MES: Multiple Event Statistic in \kepler jargon). 

We remark that Eqs.~(\ref{eqn:sprcrit}) and~(\ref{eqn:significance_true_planet}) are, respectively, corrected versions of Eqs.~(28) and~(23) in \cite{2019Marchioripaper}. The corrections address the proper placement of the $(1 - \rm SPR_{tot})$ factor to ensure mathematically consistent flux ratio treatments between target and total flux measurements. In more detail, the proper treatment of target flux ($\rm F_{T}$) to total flux ($\rm F_{tot}$) ratios requires $\rm F_T/F_{tot} = \rm 1 - SPR_{tot}$, which was incorrectly applied in the original formulations.   

\section{Double-aperture photometry and flux measurements for detecting FPs}
\label{sec:Double_Aperture}
The idea of the  double-aperture photometry  is to use an extra aperture to compute an alternative photometry to the nominal photometry in order to detect FPs. Here we consider two versions of double-aperture photometry: the extended mask and the secondary mask. Sections~\ref{subsec:ext} and~\ref{subsec:sec} respectively present and describe each one of these masks. Sect.~\ref{sec:Efficiency} explains that extended and secondary masks can also be used alongside centroid shift measurements. This allows to create the extended centroids when using extended masks as well as the secondary centroids when using secondary masks. In Sect. \ref{sec:Conclusions} we draw a scheme to indicate different applicable scenarios for each mask. 

\subsection{Extended mask and extended flux}
\label{subsec:ext}
The extended mask is an expansion of the nominal mask. Given the scenario of a target surrounded by contaminants, collecting their flux is useful to look for FPs. In order to collect that contaminant flux we increase the size of the nominal mask. This is done to reach as much contaminants as possible. The expansion process is done by surrounding the nominal mask by a given number of pixels, typically one pixel, in order to avoid increasing the noise significantly. Fig \ref{fig:extended} shows an schematic representation of how an extended mask is built. 

At window level, there will be an overlapping between the nominal and the extended masks. A large fraction of the flux collected by the extended mask comes from the target. However, the extended mask collect a larger fraction of the flux from the contaminant stars compared to the nominal mask, and hence it is more sensitive to the existence of a transit on a contaminant star. From this perspective the extended mask can be seen as a nominal mask that is more sensitive to all the contaminant flux in the imagette. If a transit happens in a contaminant star of index $\rm k$, the observed transit depth in the extended mask is called $\rm \delta_{k}^{ext}$ and is
\begin{equation}
     \rm \delta_{k}^{ext} = \delta_{\rm EB} \, SPR_{k}^{ext}\;,
     \label{eqn:delta_k_ext}
\end{equation}
where $\rm SPR_{k}^{ext}$ is computed with Eq.~(\ref{eqn:spr_k}) by using the extended mask. We can define the statistical significance in the extended mask for a transit happening in a contaminant star
\begin{equation}
    \rm \eta_{k}^{ext} =  \frac{ \deltaEB SPR_{k}^{ext}  \sqrt{\rm  \teb \, n_{tr}}}{(1 - SPR_{tot}^{ext})NSR_{1h}^{ext}}\;, 
\end{equation}
where $\rm SPR_{tot}^{ext}$ and $\rm NSR_{1h}^{ext}$ are, respectively, Eqs. (\ref{eqn:NSR_11}) and (\ref{eqn:sprt_tot}) assuming an extended mask. Since the photometry extraction for each mask is performed separately, two light curves will be created per window: one based on the nominal mask and the other on the extended mask. The light curve produced with the extended mask is used to detect FPs since the extended mask might enhance any transit signal coming from the contaminants. For instance, if we have a transit signal in the nominal light curve but an even deeper transit signal in the extended light curve, we can conclude that the origin of the signal is not the target but it is due to one of the contaminant stars located nearby the target. In other words we are dealing with a FP that thanks to the extended mask we can identify as such. The physical explanation for this relies on dilution effects: if a genuine planet transits the target star, the extended light curve will exhibit equal or shallower transit depth than the nominal curve due to additional flux from nearby stars in the larger aperture. Extended transits that are deeper than nominal transits are therefore physically incompatible with target-star planets and can only arise from contaminating EBs.

Of course, by increasing the size of the mask we are also increasing the NSR (lowering the SNR) of the photometric signature of the background transit, which means that a trade-off has to be established. For instance, it would be enough to increase the size of the extended mask by one more pixel and see the impact on the efficiency. This has been done already by~\cite{gutierrezcanales2025}, where Fig.~7.1 shows that creating extended masks by extending nominal masks beyond the single-pixel ring significantly reduces flux efficiency, likely because the added pixels increase noise more than signal. Therefore, even if more optimal ways to build extended masks could be implemented (on-going work in the mission consortium), like adding only specific pixels to the nominal mask (see Sect.~\ref{subsec:concluding_remarks}), we adopt for the sake of simplicity the 1-pixel ring approach for the rest of this work.

\subsection{Secondary mask and secondary flux}
\label{subsec:sec}
\begin{figure*}
     \centering
     \begin{subfigure}[b]{0.31\textwidth}
         \centering
         \includegraphics[width=\textwidth]{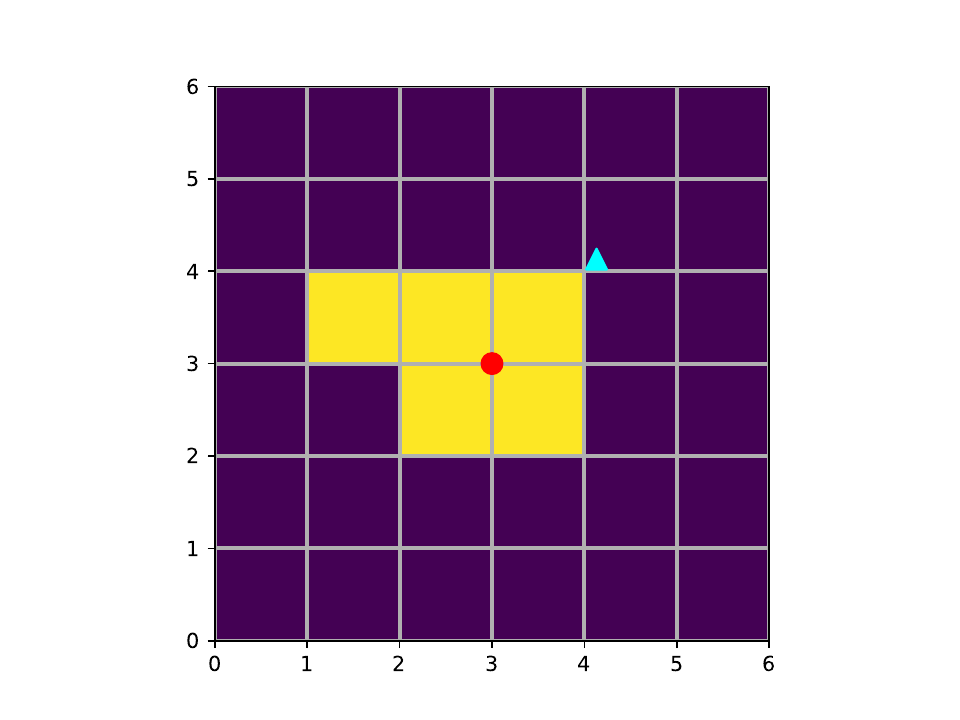}
         \caption{Nominal mask}
         \label{fig:nominal}
     \end{subfigure}\hspace{0.01\textwidth}
     \begin{subfigure}[b]{0.31\textwidth}
         \centering
         \includegraphics[width=\textwidth]{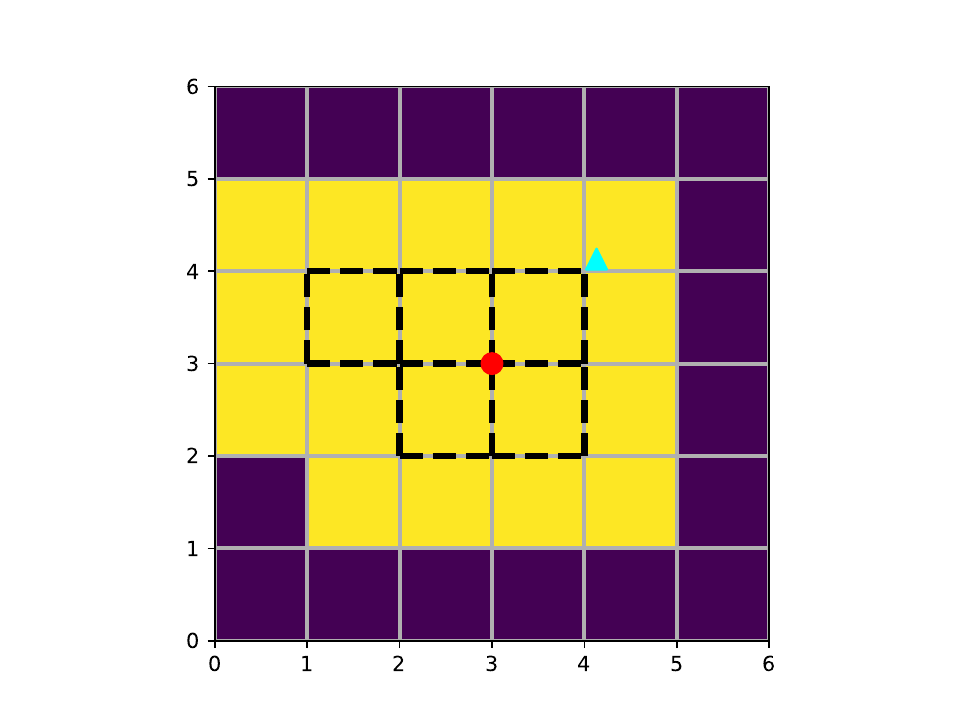}
         \caption{Extended mask}
         \label{fig:extended}
     \end{subfigure}\hspace{0.01\textwidth}
     \begin{subfigure}[b]{0.31\textwidth}
         \centering
         \includegraphics[width=\textwidth]{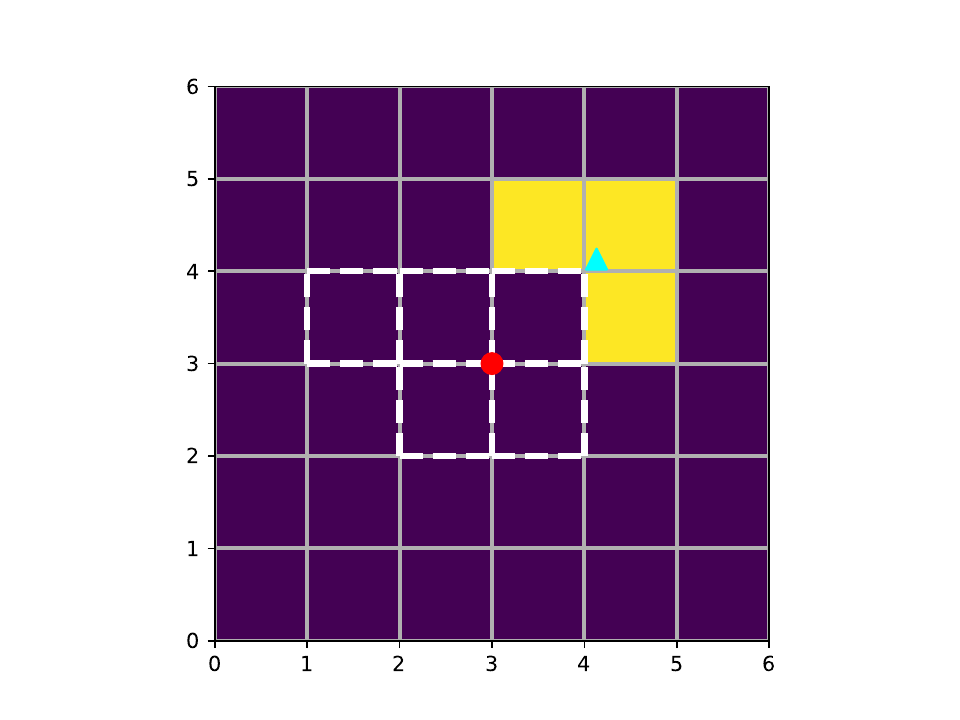}
         \caption{Secondary mask}
         \label{fig:secondary}
     \end{subfigure}
        \caption{Schematic view (from left to right) of example nominal, extended and secondary masks. For the extended and secondary masks, the respective nominal mask is represented with dashed lines. In this case the nominal mask consists in 5 pixels, the extended mask in 19 pixels and the secondary mask in 3 pixels. The red circle represents the target star and the  cyan triangle represents a nearby contaminant star that could be an EB. Nominal masks are used to extract target photometry and are built to have the lowest NSR. Extended masks are extensions of nominal masks, such as each nominal mask is surrounded by a ring of usually one pixel. Secondary masks are typically smaller and are centered on the most problematic contaminant star in the window.}
        \label{fig:three masks}
\end{figure*}

The idea behind the concept of a secondary mask is to collect the flux only from one contaminant in each window. That contaminant is the one with the highest $\rm SPR_{k}$ value. This is the reason we sometimes refer to that contaminant as the most prominent. The reason behind this choice is that the contaminant with the highest $\rm SPR_{k}$ value is also the most probable to cause a FP. However, the observed transit depth of a transit happening in the most prominent contaminant star in the secondary mask, $\rm \delta_{kmax}^{sec}$, can be diluted by nearby contaminants. This is expressed as
\begin{equation}
    \delta_{\rm kmax}^{\rm sec} =  \delta_{\rm EB}  \, \rm SPR_{\rm  kmax}^{\rm  sec}\;,
    \label{eqn:delta_kmax_sec}
\end{equation} 
where $\rm SPR_{k}^{sec}$ is computed with Eq.~(\ref{eqn:spr_k}) by using the secondary mask.
Then, as done for the nominal mask (see Sect.~\ref{subsec:detectability_of_transit_signals}),  we derive the  statistical significance in the secondary mask for a transit happening in the most prominent contaminant star
\begin{equation}
    \rm \eta_{kmax}^{sec} = \frac{\deltaEB  \sqrt{ \teb \ntr}}{NSR_{1h}^{sec}}\;,
    \label{eqn:eta_sec}
\end{equation}
where $\rm NSR_{1h}^{sec}$ refers to Eq. (\ref{eqn:NSR_11}) assuming a secondary mask and the most prominent contaminant in the window having the role of the target. This means that $\rm I_{n}^{kmax}$ appears in the denominator of $\rm NSR_{1h}^{sec}$. 
The secondary mask can be seen as a nominal mask that is centered not in the target but in the contaminant with the highest value of $\rm SPR_{\rm k}$ in the window.  This  is the reason Eq.~(\ref{eqn:eta_sec}) is formally equivalent to Eq.~(\ref{eqn:significance_true_planet}), which corresponds to the significance of the planet transiting the target star. From this perspective,  the process to obtain the secondary mask is analogous to the one for obtaining the nominal mask (described in Appendix \ref{sec:Apendix_B}).  Given the fact that most contaminants are fainter than the target, the resulting secondary mask is typically only a few pixels in size. This is why secondary masks have --~ in most cases ~-- a lower NSR (high SNR) than the extended masks. Fig~\ref{fig:secondary} shows an example view of the secondary mask.

Like for the extended mask, two light curves will be produced for every window containing a nominal and a secondary mask. This "secondary" light curve can be used to detect FPs in the following way: if we have a transit signal in the nominal mask but an even deeper transit  in the secondary mask, we can conclude that the origin of the signal in the nominal mask was in fact the star where the secondary mask is centered. From this perspective, the secondary mask works as a focused nominal mask that enhances the transit signal from a single, specific contaminant, instead of dealing with several contaminants at the same time, as in the case of the extended mask. However, this can also be seen as a possible drawback for secondary masks, specially for very crowded windows. If more than one contaminant can create a FP in a window, monitoring as much contaminants as possible is crucial. In Sect.\ref{subsec:overall_efficiency_of_the_different_metrics} we explore this in more detail.

\section{Centroid shifts for detecting FPs}
\label{sec:Centroid}
The centroid shift method is based on detecting a shift in the center of brightness of an window whenever there is a transit. We refer as to the centroid (often called ``Center of Brightness" or ``Photocenter") of a given window to the coordinates of the geometric center of all the flux sources in the window. Centroid measurements are computed using only pixels within a given aperture. We recall now that --~ for the targets for which photometry is extracted on-board ~-- the current strategy for \plato to detect FPs includes the computation of centroid shifts using the nominal mask. However, just a small amount of P5 targets will have those measurements.

When no transits occur, the centroid of the window remains close to the target. If there is a transit on the target, a small centroid shift is created. If a transit happens on a contaminant star that is bright and distant enough, but not too far away, from the target, the resulting centroid is enhanced. As pointed out by \citet{2017Gunther}, if the system (target + contaminants) can be visually resolved, the direction of the shift indicates which object undergoes the eclipse.

The coordinates of the centroid in the window reference frame are 
\begin{equation}
\rm C_{x} = \frac{\sum\limits_{n=1}^{36} \rm x_{n} \rm I_{n} \omega_{n}}{F_{\rm tot}},  \; \rm C_{y} = \frac{\sum\limits_{n=1}^{36}  \rm y_{n} \rm I_{n} \omega_{n}}{F_{\rm tot}} \; ,
\label{eqn:x-y-centroid}
\end{equation}
where $\rm F_{\rm tot}$ is given in Eq.~(\ref{eqn:F_tot}), $\rm I_{n}$ is the total intensity in the pixel of order $\rm n$ (Eq.~(\ref{eqn:I_n})), $\rm x_{n}$ is the row coordinate of the center of every pixel in the window frame, $\rm y_{n}$ is consequently the column coordinate of the center of every pixel in the window frame and $\omega_{n}$ is the considered mask.

When a background transit occurs,  $\rm I_{n}$ in Eq. (\ref{eqn:x-y-centroid}) is replaced by $\rm I_{n}^{in}$, which is the total intensity in  the pixel of order $\rm n$ during the transit. Any centroid shift is always associated with a given contaminant, $\rm k$, which we suppose is an EB. The corresponding $\rm x$ and $\rm y$ coordinates of the centroid during the transit are called, respectively, $\rm C_{x}^{k, in}$ and $\rm C_{y}^{k, in}$. Note that here the centroid shift in the transit is indexed with $k$ since it depends on the contaminant of index $k$ at the origin of the background transit. 

We can define now the centroid shift in each direction. The change in the position of the centroid alongside the $\rm x$ direction is $\rm \Delta C_{x}^{k} = C_{x} - C_{x}^{k, in}$ and the change alongside the $\rm y $ direction is $\rm \Delta C_{y}^{k} = C_{y} - C_{y}^{k, in}$. The absolute (or total) centroid shift is
\begin{equation}
    \rm \Delta C^{k} = \sqrt{ \rm \left(\Delta C_{\rm x}^{k}\right)^{2} + \left( \Delta C_{\rm y}^{k} \right)^{2}}\;.
\end{equation}
We can as well define the centroid shift alongside one direction in terms of transit depth as follow
\begin{equation}
    \rm \Delta C_{x}^{k} = \lambda_k[\delta] \Gamma_{x}^{k} \; ,
\end{equation}
where
\begin{equation}
    \rm \lambda_k [\delta_{\rm EB} ] = \frac{ \delta_{\rm EB}  \times 10^{-6}}{1 -  \delta_{\rm EB}  \times 10^{-6} SPR_{k}}\; , 
    \label{eqn:lambda}
\end{equation}
and
\begin{equation}
    \rm \Gamma_{x}^{k} = \frac{\sum\limits_{n=1}^{36} (x_{n} - C_{x}) \omega_{n} I_{n}^{k}}{F_{tot}} \; .
\end{equation}
We recall that in this work the transit depths are assumed to be in parts per million (ppm). This requires conversion to dimensionless fractions $(\times 10^{-6})$ like the one in Eq.~(\ref{eqn:lambda}). Now, the absolute centroid shift can be rewritten as
\begin{equation}
    \rm \Delta C_{k}  = \lambda_k[\delta_{\rm EB} ] \sqrt{(\Gamma_{\rm x}^{\rm k} )^{2} + (\Gamma_{\rm y}^{\rm k})^{2}} = \lambda_k[\delta_{\rm EB} ] \Gamma_{k} \; ,
    \label{eqn:delta_gamma_centroid_shift}
\end{equation}
where we have defined
\begin{equation}
    \rm \Gamma_{k} = \sqrt{ (\Gamma_{\rm x}^{\rm k})^{2} + (\Gamma_{\rm y}^{\rm k})^{2}  } \; .
\end{equation}

\subsection{Centroid shift error}
\label{subsec:Centroid_error}
In order to compute the error for the absolute centroid shift we need to compute first the error for the shift in each direction. We introduce the term $\delta \rm I_{n} = I_{n} - \bar{I}_{n}$ as an additional help in our calculations. The term $\rm \bar{I}_{n}$ refers to the mean of the variable $\rm I_{n}$, which is  given by Eq.~(\ref{eqn:I_n}). Now we can write the variances as
\begin{equation}
        \rm Var(\delta I_{n})  =  \bar{I}_{n}  + {\rm B} \, \Delta t_{\rm exp} \, + \sigma_{D}^{2} + \sigma_{Q}^{2}\;, \\
\end{equation}
where $\rm B$, $\Delta \rm t_{\rm exp}$,  $\rm \sigma_{D}^{2}$, $\sigma_{Q}^{2}$ are the quantities involved  in Eq.~(\ref{eqn:NSR_11}) and their values are given in Table~\ref{tab:symbols}.
With this new term we can compute the noise associated to the centroid on the $\rm x$ direction, $\sigma_{\rm x}^{2}$, as follows (the noise for the centroid on the $\rm y $ direction is analogous) 
\begin{equation}
    \sigma_{\rm x}^{2} =  \frac{\sum\limits_{n=1}^{36} (\rm x_{\rm n} - \rm C_{\rm x})^2 \omega_n {\rm Var} \left( \delta I_{\rm n} \right)}{\rm F_{\rm tot}^{\rm 2}} \;.
\label{eqn:sigma_x}
\end{equation}
Eq. (\ref{eqn:sigma_x}) is actually a revision of Eq.~(2) from \cite{2013PASP..125..889B}, which is not invariant under translations of the reference system. 

The error for the absolute centroid shift is therefore  
\begin{equation}
    \sigma_{ \Delta_{\rm C}} = \frac{1}{\Delta \rm C} \sqrt{\Delta \rm C_{x}^{2} \sigma_{\rm x}^{2} + \Delta \rm C_{y}^{2} \sigma_{\rm y}^{2}} 
    \label{eqn:sigma_centroid} \;,
\end{equation}
where we have assumed a larger number of out-of-transit centroid measurements than for the in-transit ones. This gives more precision for the out-of-transit centroids than for the in-transit centroids. {Accordingly, we can assume $\rm \sigma_{\rm x}^{\rm out} << \sigma_{\rm x}^{\rm in}$ (and analogously $\rm \sigma_{\rm y}^{\rm out} << \sigma_{\rm y}^{\rm in}$ ).

Furthermore,  Eq.~(\ref{eqn:sigma_centroid}) is  equivalent to Eq.~(7) from \cite{2013PASP..125..889B}. There is however a difference in the reference system of both papers. In this work the reference system for the centroid computations is the CCD reference frame while \cite{2013PASP..125..889B} use equatorial coordinates (R.A. and declination).

To derive Eq.(\ref{eqn:sigma_centroid}) we have assumed that the centroid error is always significantly smaller than the centroid shift itself. As can be seen, Eq.~(\ref{eqn:sigma_centroid}) goes to zero as the centroid shift, $\rm \Delta C$, goes to zero.  This is however problematic for scenarios where the centroid shift is very small and comparable to the centroid error. One of these scenarios is the one where centroid shifts are computed with secondary masks of only one pixel in size. For such cases centroid shifts are naturally zero and therefore Eq.(\ref{eqn:sigma_centroid}) no longer holds. More details about how we avoided such cases are given in Sects.~\ref{sub_subsec:nominal_centroid_shift}, \ref{sub_subsec:extended_centroid_shift}  and~\ref{sub_subsec:secondary_centroid_shift}.

\subsection{Significance of the Centroid}
\label{subsec:significance_of_the_centroid}
We can define the statistical significance of a centroid shift signal taking into account some PLATO specifications and scientific requirements. For instance we can average centroid measurements over a duration of one hour and a given number of cameras, $\rm N_{T}$, in order to accordingly reduce the uncertainty as follows:
\begin{equation}
    \sigma^{\rm 1h, N_{\rm T}} = \frac{\sigma_{\Delta_{\rm C}}}{12 \sqrt{\rm N_{\rm T}}} \; .
    \label{eqn:centroid-noise}
\end{equation}
We recall that, in a parallel way to flux measurements described in Sect.~\ref{subsec:detectability_of_transit_signals}, we compute the centroid uncertainty after averaging the centroid over one hour.

The corresponding statistical significance of the centroid shift in the nominal mask for a transit happening in a contaminant star is
\begin{equation}
    \eta^{\rm nom, \Delta C}_{k} = \rm \frac{ \Delta C_{k} \sqrt{  \teb  \ntr}}{ \sigma^{\rm 1h, N_{\rm T}}} = \rm \frac{ \lambda_{\rm k}[ \delta_{\rm EB} ]\Gamma_{\rm k}^{\rm nom} \sqrt{ \teb \ntr}}{\sigma^{1h, N_{\rm T}}} \; .
    \label{eqn:eta_centroid_shift}
\end{equation}
While the significance of the background transit in the flux scales as $\deltaEB \sqrt{\teb}$, the statistical significance of the centroid shift rather scales as $\rm \lambda [\deltaEB] \sqrt{\teb}$. However, the denominator in Eq.~(\ref{eqn:lambda}) is for most contaminant stars close to 1 such that in most cases $\lambda  \propto \delta_{k}$. Accordingly,  we can consider that as for the significance in terms of photometry (Eq.~(\ref{eqn:bt_eta})), the significance of the centroid shift predominantly scales as $\rm \deltaEB \sqrt{\teb}$.

The subscript ``nom" in Eq.~(\ref{eqn:eta_centroid_shift}) indicates that we are using the nominal mask. However,  we can replace the nominal mask with either the extended or secondary masks and have equivalent expressions as follows 
\begin{eqnarray}
        \eta^{\rm ext, \Delta C}_{k} & = &  \rm \frac{ \lambda_k[ \deltaEB]\Gamma_{k}^{\rm ext} \sqrt{ \teb \ntr}}{\sigma_{\rm ext}^{\rm 1h, N_{\rm T}}} \;
        \label{eqn:eta_extended_centroid_shift}\\
        \eta^{\rm sec, \Delta C}_{k}&  = &  \rm \frac{ \lambda_k[ \deltaEB]\Gamma_{k}^{\rm sec} \sqrt{ \teb \ntr}}{\sigma_{\rm sec}^{\rm 1h, N_{\rm T}}} \;
        \label{eqn:eta_secondary_centroid_shift},
\end{eqnarray}
where $\rm \sigma_{\rm ext}^{\rm 1h, N_{\rm T}} $ and $\rm \sigma_{\rm sec}^{\rm 1h, N_{\rm T}} $ refers to Eq.~(\ref{eqn:centroid-noise}) but using respectively the extended and secondary masks instead.

\section{Methods and assumptions}
\label{sec:methods}
We present now our methodology and key assumptions, particularly regarding contaminant stars and EBs. We recall that our goal is to see how efficient are flux and centroid shift measurements when using double-aperture photometry for detecting FPs. For this goal, we compared the flux and centroid measurements  based on our alternative apertures (extended and secondary masks) with the  centroid shifts computed with the nominal mask. 

\subsection{Stellar sample}
\label{sec:stellar_sample}
We used a stellar sample with data from Gaia DR3 \citep{2016Gaia, 2023A&A...674A...1G}. Our sample contains 13.6 millions of stars, ranging from magnitude 2.1 up to 21 in the Gaia band. This magnitude range and cutoff for faint targets comes from Gaia DR3,  rather than \plato sensitivity constraints. The sample was selected based on an approximation of the number of stars in  the FoV of a single \plato N-CAM for a single pointing at the center of the SPF (Southern \plato Field). We however make notice that this only somewhat representative of what the true respective \plato sample would look like. For simplicity, we did not take into account the proper motion of any star in the Gaia catalog. We converted the magnitude of the stars in the sample from the Gaia band to the \plato magnitude system by using Eq.~(9) from \cite{2019Marchioripaper}, where $\rm P$ is related to $\rm G$, $\rm G_{\rm bp}$ and $\rm G_{\rm rp}$ magnitudes measured by Gaia. Afterwards, and taking into account \plato P5 sample specifications described in Sect. \ref{sec:PLATO_inst}, we defined as a target every star with magnitude between 8.0 $\leq$ P $\leq$ 13.0. We defined as contaminant stars any star in the whole magnitude range of the sample, within a distance of 10 pixels for every target. This since the probability of having a FP generated by a contaminant located at a distance larger than 10 pixels is very low according to \cite{2019Marchioripaper}.

\subsection{Assumptions about background Eclipsing Binaries}
\label{subsec:assumptions}

In principle, the information that \plato will have about contaminant stars comes from Gaia and the PIC (\plato Input Catalog) \citep[see][]{PLATO_PIC}, where a flag system for EBs can be implemented. However, there are studies about EB populations like \cite{Soederhjelm_Dischler_2005} and \cite{2023Mowali}. This last one is of particular interest since they report approximately two million EB candidates identified with Gaia. In terms of EB occurrence rates for space missions, we can mention \cite{Prsa_2011} that reported and EB occurrence of 1.2\,\% for \kepler\footnote{This estimation is $\sim 50\,\%$ higher than another, previous one produced with Hipparcos data. The authors suggest that the increase is due to \kepler's photometric superiority.} and \citet{Prsa_2022} that reported an EB occurrence of 3\,\% at low Galactic latitudes for \tess. However, even if we recognize it is not feasible to know all contamination sources for each individual window or imagette for \plato in advance, we don't aim here to estimate realistic EB rates for \plato, as we mentioned in the introduction, and we refer to the work by \citet{2023MNRAS.518.3637B} on that regard.
Another important assumption of this work is related to stellar variability. We did not include in our flux or centroid shift calculations intrinsic stellar variability out for simplicity. 

Following \citet{2019Marchioripaper}, we assumed that all contaminants in this work are EBs but instead of assuming that all EBs have the same transit depth and duration, we refined their approach by drawing their properties from known distributions. For each EB we randomly sampled $(\deltaEB, \teb)$ pairs from the Kepler Eclipsing Binary Catalog \citep{2016AJ....151...68K} \footnote{\url{http://keplerebs.villanova.edu/}} and transit duration distribution in the Certified FP Table \footnote{\url{https://exoplanetarchive.ipac.caltech.edu/docs/data.html}}. However, we note that assuming all contaminants are EBs represents a worst-case scenario for contamination. This allow us to assess the maximum capability of each detection method under challenging conditions.

\subsection{Instrument performance and PSF assumptions}
The Point-Spread Function (PSF) describes how a stellar signal is distributed after interacting with the optics of an instrument. In the case of \plato, the PSF determined how the stellar signal is distributed over the pixels of the detectors. We used 4224 \plato-representative PSFs that were created in ZEMAX by simulating the baseline optical layout of the instrument. These PSFs also include miss-alignment and mounting errors. They were computed for the \plato reference 6000 K G0V star and for the best focus (0~$\mu$ m distance to the best focus).

Furthermore, a convolution with a Gaussian kernel with a given width was applied on each PSF to mimic the charge diffusion effect in \plato CCDs. Two widths were adopted, 0.1~px and 0.2~px. The difference between the results produced with each width is marginal. At the end we adopted the PSFs convoluted with a 0.2~px width kernel. We made this decision because CCD diffusion strongly depends on the effective temperature of an observed star and a 0.2~px width approximates the aforementioned case of \plato reference stellar source: a 6000~$\rm K$ GOV star.

Once we obtained the PSFs, we looked for their coordinates in the focal plane to find the closest one to each target star. After this process, we decomposed every PSF into b-spline polynomials that are analytically integrable. This is done to accurately integrate the PSF over a given  array of pixels, in our case, a 6~$\times$~6~pixel window. For this work we assume that each camera has the same set of PSFs. Each  PSF is however not constant as it varies across the whole FoV. Furthermore, our PSF set reflects realistic camera-to-camera variations based on Monte Carlo simulations of alignment and mounting errors. These simulations were carried out by a member of the \plato Mission Consortium (PMC) and distributed within the consortium. Fig.  \ref{fig:psfs} shows some example PSFs used in our work. The PSFs are shown at three different angles of the FoV of one camera, before and after adding the diffusion effect. The corresponding windows for each PSF are shown in the last column of the figure.
\begin{figure*}[htbp]
    \centering
    \includegraphics[width=0.85\textwidth]{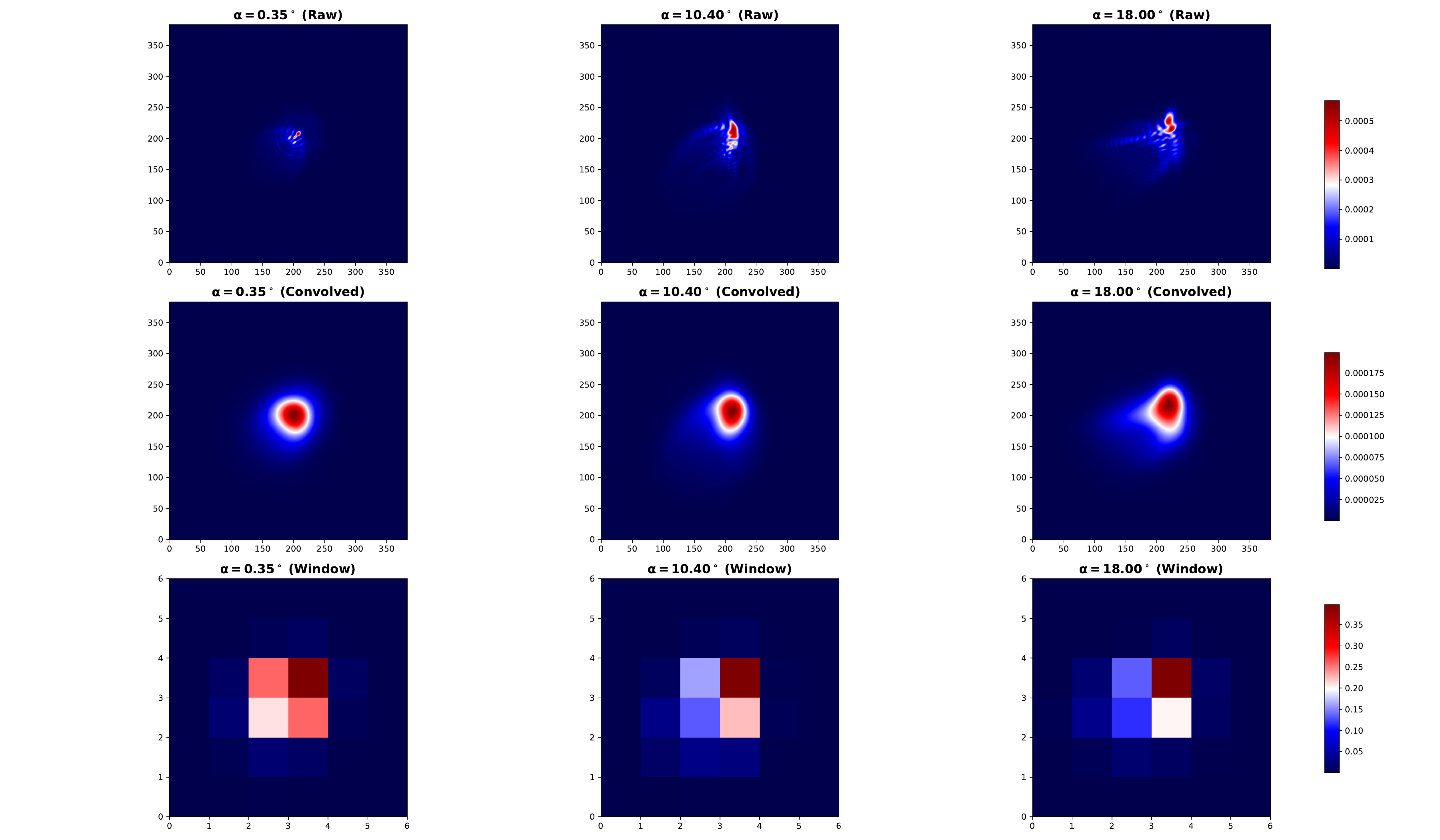}
    \caption{Simulated \plato PSFs (1/64 pixel resolution) at different angular positions, $\rm \alpha$, of a flux source in the FoV of a \plato camera. When $\rm \alpha = 0^{\circ}$ the source is in the center of the FoV and when $\rm \alpha = 18^{\circ}$ the source is at the edge of the FoV. The top row shows the PSFs without charge diffusion while on the middle row a Gaussian kernel was applied to simulate charge diffusion in the CCD. The bottom row shows the PSF integrated over the pixels of a 6x6 window. The color scheme goes from the highest value (red) to the smallest values (blue).}
    \label{fig:psfs}
\end{figure*}
The \plato cameras are designed to enclose about 77 $\%$ of the PSF flux within a 2.5 pixel window and 99 $\%$  of the flux within a 5x5 pixel window. We also notice that photometry performance depends on the knowledge of the PSF and the shape of the PSF is expected to change after launch. This implies that the 'true' PSF cannot be obtained. In order to reconstruct the 'true' PSF a microscanning technique will be implemented on-board \citep[see][]{2019SamadiPSLS}. 

\section{Efficiency of flux and centroid shift measurements for detecting FPs}
\label{sec:Efficiency}

\subsection{Efficiency of flux measurements}
\label{subsec:efficiency_double_aperture_photometry}
\subsubsection{Efficiency of extended flux}
\label{sub_subsec:extended_mask}

The efficiency of extended flux measurements requires two quantities. The first quantity is the total number of FPs expected for the nominal mask, which we call $\rm N_{FP}$. This is given by the number of contaminants per window for which Eq. (\ref{eqn:falsepos_etas}) holds ($\rm \eta_{k}^{nom} > \eta_{min}$). This is
\begin{equation}
    \rm N_{FP} = \sum\limits_{targets} \left( \sum\limits_{k=1}^{N_{C}=10}\eta_{k}^{nom} > \eta_{min}  \right) \; .
    \label{eqn:nfp}
\end{equation}
We recall that $\rm \eta_{min} = 7.1$. To compute $\rm N_{FP}$ we have to obtain the value of $\rm \eta_{k}^{nom}$ for every contaminant in the window and then compare it to $\rm \eta_{min}$. However, the number of contaminants per window, $\rm N_{c}$, can be up to hundreds. For simplicity, we computed $\rm \eta_{k}^{nom}$ for only the 10 most significant contaminant stars in the window, based on their $\rm SPR_{k}$ values. Section~\ref{sec:Results} justifies this approach, as only a few contaminants per window can induce a FP.

The second quantity for computing the efficiency of the extended mask is the number of detectable FPs by the extended mask, which we call $\rm N_{FP}^{ext}$. To compute $\rm N_{FP}^{ext}$ we count the number of times the following three conditions are fulfilled: The observed transit depth in the extended mask, $\rm \delta_{k}^{ext}$, significantly exceeds that in the nominal mask; the statistical significance in the extended mask, $\rm \eta_{k}^{ext}$, surpasses the threshold that we call $\rm \eta_{min}^{ext} $ (which value is 3), and finally the transit event has to be detectable in the nominal mask. All this can be summarized as follow 
\begin{equation}
    \rm cond_{\rm k}^{ext} = (\delta_{k}^{ext} > \delta_{k}^{nom} + 3 \, \sigma_\delta) \, \&  \, (\eta_{k}^{ext} > \eta_{min}^{ext}) \, \&  \, (\eta_{k}^{nom} > \eta_{min}) \; ,
    \label{eqn:cond_ext_raw}
\end{equation}
where we have defined
\begin{equation}
\sigma_{\rm \delta} = \sqrt{ \sigma_{\rm \delta,ext}^2 +\sigma_{\rm \delta,nom}^2 } \; ,
\label{eqn:sigma_delta}
\end{equation}
where $\sigma_{\rm \delta,ext}$ (resp. $\sigma_{\rm \delta,nom}$) is the 1-$\sigma$ uncertainty associated with the transit depth measurement with the extended flux (resp. nominal flux). 
The condition $(\rm \delta_{k}^{ext} > \delta_{k}^{nom} + 3 \, \sigma_\delta)$ in Eq.~(\ref{eqn:cond_ext_raw}) ensures that the (apparent) transit depth measured in the extended flux is significantly larger than in the nominal flux. It must be pointed out that in Eq.~(\ref{eqn:sigma_delta}) we are assuming both analysis of the extended and nominal flux measurements are done independently and hence the noise associated with each transit depth measurement adds quadratically.  In fact extended and nominal masks have a non-negligible number of pixels in common. This means the noises associated to the transit depths of each mask are not strictly independent. However, in this work we assumed two independent measurements of the apparent transit depths. A more in-depth approach is considering a differential analysis of both extended and nominal light curves. By differential analysis we mean to study simultaneously the extended and nominal light curves. To do this, we define the concept of ``differential transit depth''. This concept refers to the difference $ \rm \Delta \delta(t) = \rm \delta_{\rm k}^{\rm ext}(t) - \delta_{\rm k}^{\rm nom}(t)$. The differential analysis consists in measuring $\rm \Delta \delta(t)$. This analysis can substantially reduce the noises and make the detection of FP  more sensitive (see discussion in Sect.~ \ref{subsec:overall_efficiency_of_the_different_metrics} ).

By definition of the significance, 
$\rm \eta_k^{nom} = 1$ is reached when the (apparent) transit depth equals the \onesigma  dispersion in the light curve. 
Accordingly, by substituting in Eq.~(\ref{eqn:bt_eta}) the apparent transit depth $\rm \delta_{\rm k} \,  SPR_{k} $ by $\sigma_{\rm \delta,nom}$ and by imposing $\rm \eta_k^{nom} = 1$, one derives 
 \begin{equation}
\sigma_{\rm \delta,nom}   =  { { \nsr \, \left ( 1 - \sprtot \right ) } / {\sqrt{ \teb \, \ntr }} } \, .
\label{eqn:sigma_delta_nom}
\end{equation}
For the extended mask, for the same reason we have 
 \begin{equation}
\sigma_{\rm \delta,ext}  =  { { \nsrext \, \left ( 1 - \sprtotext \right ) }  / {\sqrt{ \teb \, \ntr }} }  \, .
\label{eqn:sigma_delta_ext}
\end{equation}
As for the centroid error (see Sect.~\ref{subsec:Centroid_error}), these estimates assume a larger number of out-of-transit flux measurements than for the in-transit ones. Now, the condition $(\rm \delta_{\rm k}^{\rm ext} > \delta_{\rm k}^{\rm nom} + 3 \, \sigma_\delta)$  in Eq.~(\ref{eqn:cond_ext_raw}) can equivalently be rewritten in a dimensionless form as
 \begin{equation}
\sprkext > \sprk + 3 \, \rm SPR_{\rm t} \; ,
\end{equation}
where we have defined
\begin{eqnarray}
\rm SPR_{\rm t} & =  &  \sqrt{\rm SPR_{\rm t,ext}^2 + \rm SPR_{\rm t,nom}^2}\;,  \label{eqn:SPR_t} \\
\rm SPR_{\rm t,ext} & = & \frac{ \nsrext \, (1 - \sprtotext)}{ \, \deltaEB  \, \sqrt{ \teb \, \ntr}}  \label{eqn:SPR_t_ext}\;, \\
\rm SPR_{\rm t,nom} & =  & \frac{\rm NSR_{\rm 1h} \, (1 - \sprtot)}{ \deltaEB   \, \sqrt{ \teb \, \ntr}} \, .  \label{eqn:SPR_t_nom}
\end{eqnarray}
The term $\gamma$ defined by Eq.~(19) has been introduced for the same reason than for the significance (see Eq.~(\ref{eqn:bt_eta}) and Sect.~\ref{subsec:assumptions}).
Accordingly, the condition of Eq.~(\ref{eqn:cond_ext_raw}) can be rewritten as
\begin{eqnarray}
    \rm cond_{\rm k}^{\rm ext} & = & (\sprkext > \sprk + 3 \, \rm SPR_{\rm t}) \, \&   \, \nonumber \\ 
    & & (\eta_{\rm k}^{\rm ext} > \eta_{\rm min}^{\rm ext}) \, \&  \, (\eta_{\rm k}^{\rm nom} > \etamin ) \; .
    \label{eqn:cond_ext}
\end{eqnarray}

Then, the total number of (background) transit detected by the extended flux is 
\begin{equation}
    \rm N_{FP}^{ext} = \rm \sum_{targets} \left( \sum\limits_{k=1}^{10} cond_{k}^{ext} \right)  \;.
\end{equation}
The efficiency of the extended flux measurements is the ratio of $\rm N_{FP}^{ext}$ to $\rm N_{FP}$:
\begin{equation}
    \rm Eff_{ext} = \frac{N_{FP}^{ext}}{N_{FP}} \;.
    \label{eqn:eff_ext_mask}
\end{equation}

\subsubsection{Efficiency of secondary flux}
\label{sub_subsection:secondary_mask}
The efficiency of secondary flux measurements requires two quantities. The first quantity is the total number of FPs the nominal mask is sensitive to. However, and unlike for the extended mask, for secondary flux measurements we consider only FPs induced by the most significant contaminant star in every window. This significant contaminant is identified as the one with the highest $\rm SPR_{k}$. We proceed this way because secondary masks are always centered only on one contaminant per window. At the same time, and to increase our chance to detect the background transit, we choose the contaminant for which the signature will be the higher, which is in the majority of cases the one with the highest $\rm SPR_{k}$. 

We count the number of times the most significant contaminant is able to create a detectable signal in the nominal mask ($\rm \eta_{kmax}^{nom} > \eta_{min}$) and call this number $\rm N_{FP}^{single}$. This is
\begin{equation}
    \rm N_{FP}^{single} = \sum\limits_{targets} \left( \eta_{kmax}^{nom} > \eta_{min} \right)\;.
    \label{eqn:nfp_sec_single}
\end{equation}
The second quantity for computing the efficiency of secondary flux measurements is the number of detectable FPs by the secondary mask.  We call this number $\rm N_{FP}^{sec}$. To compute it, in an analogous way to the extended mask, we count the number of times the following three conditions are fulfilled: The observed transit depth in the secondary mask, $\rm \delta_{kmax}^{sec}$, significantly exceeds that in the nominal mask; the statistical significance in the secondary mask, $\rm \eta_{kmax}^{sec}$, surpasses the threshold that we call $\rm \eta_{min}^{sec}$ (which value is 3), and the transit event is detectable in the nominal mask. All this is summarized as follow
\begin{eqnarray}
    \rm cond_{kmax}^{sec} & = & \left (\rm \delta_{kmax}^{sec}  > \rm \delta_{kmax}^{nom} + 3 \, \sigma_{\delta}^{kmax} \right ) \, \& \,  \nonumber \\ 
    & & (\eta_{\rm kmax}^{\rm sec} > \eta_{\rm min}^{\rm sec}) \,  \& \, (\eta_{\rm kmax}^{\rm nom}  >\eta_{\rm min})\; ,
    \label{eqn:sec_condition_sigma}
\end{eqnarray}
where we have defined
\begin{equation}
    \sigma_{\rm \delta}^{\rm kmax} = \sqrt{\sigma_{\rm \delta, nom_{\rm kmax}}^{2} + \sigma_{\rm \delta, sec}^{2}}\;,
\end{equation}
where $\rm \sigma_{\delta, nom_{\rm kmax} }$ is Eq.~(\ref{eqn:sigma_delta_nom}) for the most prominent contaminant and 
\begin{equation}
    \sigma_{\rm \delta,sec}  =  { { \nsrsec \, \sprkmaxsec  / {\sqrt{ \teb \, \ntr }} }}  \, .
\label{eqn:sigma_delta_sec}
\end{equation}
The condition $(\rm \delta_{kmax}^{sec} > \delta_{kmax}^{nom} + 3 \sigma_{\delta}^{\rm kmax})$ from Eq.~(\ref{eqn:sec_condition_sigma}) can equivalently be rewritten in a dimensionless form as

\begin{eqnarray}
    \rm cond_{kmax}^{sec} & = & \left ( \rm SPR_{\rm  kmax}^{\rm  sec}  > \rm SPR_{\rm kmax} + 3 \, SPR_{\rm s} \right ) \, \& \,  \nonumber \\ 
    & & (\eta_{\rm kmax}^{\rm sec} > \eta_{\rm min}^{\rm sec}) \,  \& \, (\eta_{\rm kmax}^{\rm nom}  >\eta_{\rm min})\; ,
    \label{eqn:sec_condition}
\end{eqnarray}
where $\rm SPR_{s}$ is given by
\begin{eqnarray}
\rm SPR_{\rm s} & =  &  \sqrt{\rm SPR_{\rm t,sec}^2 + \rm SPR_{\rm t,nom}^2}\;,  
\\
\rm SPR_{\rm t,sec} & = & \frac{\nsrsec \,  \sprkmaxsec }{\deltaEB \, \sqrt{ \teb \, \ntr}}  \label{eqn:SPR_t_sec} \, ,
\end{eqnarray}
and where $\rm SPR_{\rm t,nom}$ is given by Eq.~(\ref{eqn:SPR_t_nom}).
As for the extended mask, the condition $\rm (SPR_{kmax}^{sec} > SPR_k + 3 \, SPR_s)$ assumes that the analysis of the alternative flux (here the secondary flux) and of the nominal flux  are done independently (see  Sect.~\ref{sub_subsec:extended_mask} and Eq.~(\ref{eqn:sigma_delta})). Unlike the extended masks,  the secondary masks in general do not have pixels in common with the nominal mask. Accordingly, contrary to the extended flux, doing  a differential analysis of the secondary and nominal light curves is not going to reduce the noise and to increase the sensitivity of the secondary flux in detecting FP. Finally, $\rm N_{FP}^{sec}$ is
\begin{equation}
    \rm N_{FP}^{sec} = \rm \sum_{targets}  cond_{kmax}^{sec}\;.
    \label{eqn:nfp_sec}
\end{equation}
The efficiency of secondary flux measurements is the ratio of $\rm N_{FP}^{sec}$ to $\rm N_{FP}^{single}$:
\begin{equation}
     \rm Eff_{sec} =  \frac{N_{FP}^{sec}}{N_{FP}^{single}} \;.
    \label{eqn:eff_sec_mask}
\end{equation}
 
\subsection{Efficiency of centroid shift measurements}
The efficiency assessment of the centroid shift measurements parallels that of double-aperture photometry. We count the number of detectable FPs by centroid shifts with each mask (i.e. nominal, extended and secondary) against the total number of FPs the nominal mask is sensitive to, produced by the 10 most significant contaminants in each window. The denominator for the efficiency expressions for the centroid shifts with each mask are the same as those for double-aperture photometry. We recall now that nominal centroids are the strategy that is already envisaged for a minimal fraction of 5~$\%$ of the P5 sample. The extended and secondary centroids can be part of the additional strategy to the nominal centroids.

\subsubsection{Efficiency of nominal centroid shifts}
\label{sub_subsec:nominal_centroid_shift}
To compute the efficiency of nominal centroid shift measurements, we consider the 10 most significant contaminants (ranked by SPRk value)  within a distance of 10 pixels form the  target. First we obtain $\rm N_{FP}^{\Delta C}$, that is the count the number of detectable FPs by centroid shifts using the nominal mask. Then we count the number of times the following three conditions are fulfilled: the significance in the nominal mask of the centroid shift produced by each one of the 10 contaminants, $\rm \eta_{k}^{\Delta C}$,  exceeds a significance threshold called $\rm \eta_{min}^{\Delta C}$ (which value is 3); the transit event has to be detectable in the nominal mask and the nominal centroid shift has to be 10 times bigger than the nominal centroid uncertainty scaled by the transit duration and number of transits~(see last paragraph of Sect.~\ref{subsec:Centroid_error}). This is summarized as follow
\begin{equation}
\begin{split}
    \rm cond_{k}^{nom,\Delta C} = &(\rm \eta_{k}^{\Delta C} > \eta_{min}^{\Delta C})\, \&  \, (\eta_{k}^{nom} > \eta_{min}) \\
    & \& \rm \left(\Delta C^{nom}_{k} > 10 \frac{\sigma^{1h, N_{T}}_{nom}}{\sqrt{\teb \ntr}}\right) \; .
\end{split}
    \label{eqn:cond_nom_cob}    
\end{equation}
Then $\rm N_{FP}^{\Delta C}$ is
\begin{equation}
    \rm N_{FP}^{\Delta C} = \sum_{targets} \left( \sum\limits_{k=1}^{10} cond_{k}^{nom,\Delta C} \right) \; .
    \label{eqn:nfp_nom_centroid}
\end{equation}
The efficiency for the nominal centroid shift is
\begin{equation}
    \rm Eff_{\Delta C}^{nom} = \frac{N_{FP}^{\Delta C}}{N_{FP}} \; .
    \label{eqn:eff_nom_centroid}
\end{equation}
It is worth to note that because of the last condition in Eq.~(\ref{eqn:cond_nom_cob}), 
Eq.~(\ref{eqn:eff_nom_centroid}) has to be considered as a lower limit for 
the efficiency of the nominal centroid shift.

\subsubsection{Efficiency of extended centroid shifts}
\label{sub_subsec:extended_centroid_shift}
Similar to the nominal case, extended centroid shift efficiency is computed by considering the same 10 most significant contaminants per window. The extended mask provides enhanced sensitivity to these contaminants due to its larger size. First we obtain $\rm N_{FP}^{ext,\Delta C}$, that is the count of detectable FPs with centroid shifts using the extended mask. Then we count how many times the following three conditions are fulfilled: the significance in the extended mask of the centroid shift produced by each one of the 10 contaminants, $\rm \eta_{k}^{ext, \Delta C}$, exceeds the significance threshold $\rm \eta_{min}^{\Delta C}$ ; the transit has to be detectable in the nominal mask and the extended centroid shift has to be 10 times bigger than the extended centroid uncertainty scaled by the transit duration and number of transits. This summarized as follow
\begin{equation}
    \begin{split}
    \rm cond_{k}^{ext,\Delta C} = & \rm (\eta_{k}^{ext, \Delta C} > \eta_{min}^{\Delta C}) \, \&  \, (\eta_{k}^{nom} > \eta_{min})
    \\
    & \& \left(\rm \Delta C^{ext}_{k} > 10 \frac{\sigma^{1h, N_{T}}_{ext}}{\sqrt{\teb \ntr}}\right) \;.
    \end{split}
    \label{eqn:cond_ext_cob}
\end{equation}
Then $\rm N_{FP}^{ext,\Delta C}$ is
\begin{equation}
    \rm N_{FP}^{ext, \Delta C} = \sum_{targets} \left( \sum\limits_{k=1}^{10} cond_{k}^{ext,\Delta C} \right) \; .
\end{equation}
The efficiency of the extended centroid is
\begin{equation}
    \rm Eff_{\Delta C}^{ext} = \frac{N_{FP}^{ext,\Delta C}}{N_{FP}} \;.
    \label{eqn:eff_ext_centroid}
\end{equation}
For the same reason than the nominal centroid shift, 
Eq.~(\ref{eqn:eff_ext_centroid}) is  a lower limit for 
the efficiency of the extended centroid shift.

\subsubsection{Efficiency of secondary centroid shifts}
\label{sub_subsec:secondary_centroid_shift}
For secondary centroid shifts, we consider only the single most prominent contaminant (highest SPRk value) per window, as the secondary mask is specifically centered on this star. This represents the most problematic contaminant that is most likely to generate a FP signal. First we obtain $\rm N_{FP}^{sec,\Delta C}$, that is the count of detectable FPs with centroid shifts using the secondary mask. Then we count the times the following three conditions are fulfilled: the significance in the secondary mask of the centroid shift produced by the most significant contaminant, $\rm \eta_{kmax}^{sec, \Delta C}$,  exceeds the significance threshold  $\rm \eta_{min}^{\Delta C}$  ; the transit has to be detectable in the nominal mask and that the secondary centroid shift has to be 10 times bigger than the secondary centroid uncertainty scaled by the transit duration and number of transits. This is summarized as follow
\begin{equation}
    \begin{split}
    \rm cond_{kmax}^{sec,\Delta C} = & (\rm \eta_{kmax}^{sec, \Delta C} > \eta_{min}^{\Delta C}) \, \&  \, (\eta_{kmax}^{nom} > \eta_{\rm min}) \\ 
    & \& \left(\rm \Delta C^{sec}_{k} > 10 \frac{\sigma^{1h, N_{T}}_{sec}}{\sqrt{\teb \ntr}}\right)\;.
    \end{split}
    \label{eqn:cond_sec_centroid}
\end{equation}
As mentioned in Sect.\ref{subsec:Centroid_error} we had to account for the fact that Eq.(\ref{eqn:sigma_centroid}) no longer holds for very small centroid shifts, like the ones produced by 1-pixel secondary masks. These occurrences are  automatically discarded by the condition  $\left(\rm \Delta C^{sec}_{k} > 10 \frac{\sigma^{1h, N_{T}}_{sec}}{\sqrt{\teb \ntr}}\right)$. Then $\rm N_{FP}^{sec,\Delta C}$ is
\begin{equation}
    \rm N_{FP}^{sec,\Delta C} = \sum\limits_{targets} cond_{kmax}^{sec,\Delta C}\;.
\end{equation}
The efficiency of the secondary centroids is
\begin{equation}
    \rm Eff_{\Delta C}^{sec} = \frac{N_{FP}^{sec,\Delta C}}{N_{FP}^{single}} \;.
    \label{eqn:eff_sec_centroid}
\end{equation}
For the same reason as for the nominal centroid shift, Eq.~(\ref{eqn:eff_sec_centroid}) is  a lower limit for the efficiency of the secondary centroid shift.

\section{Results}
\label{sec:Results}

We present now the results for the scenario described in Sect.~\ref{subsec:assumptions}, analyzing how variable EB transit parameters affect FP detection efficiency and resource allocation for the metrics described in this work.

\subsection{Contamination landscape}
\label{subsec:contamination_landscape}

The metrics and procedures described on Sect.\ref{sec:Efficiency} estimates the number of FPs per window. Fig.~\ref{fig:fps} shows the distribution of $\rm N_{FP}$, Eq.(\ref{eqn:nfp}), for targets in the magnitude range $ 10\leq \rm P \leq 13 $ from the stellar sample described in Sect. \ref{sec:stellar_sample}. We select this magnitude range because the brightest P5 targets will have their photometry extracted on-ground. Our results indicate that $\sim 40\%$ of the P5 targets have no contaminant stars capable of creating a FP, while $\sim 35\%$ have only one. This suggests that for more than $70 \%$ of the P5 targets only up to one of such contaminants is able to create a FP. Additionally, $\sim 15\%$ of the targets could have two contaminant stars causing FPs. The percentage of targets that can have more than two FPs is only of $\sim 6 \%$. This is the reason behind the simplification made in Sect. \ref{sec:Efficiency}, i.e. the use of $\rm N_{\rm C} = 10$ in Eq.~(\ref{eqn:nfp}). 

\begin{figure}[htbp]
    \centering
    \includegraphics[width=0.85\columnwidth]{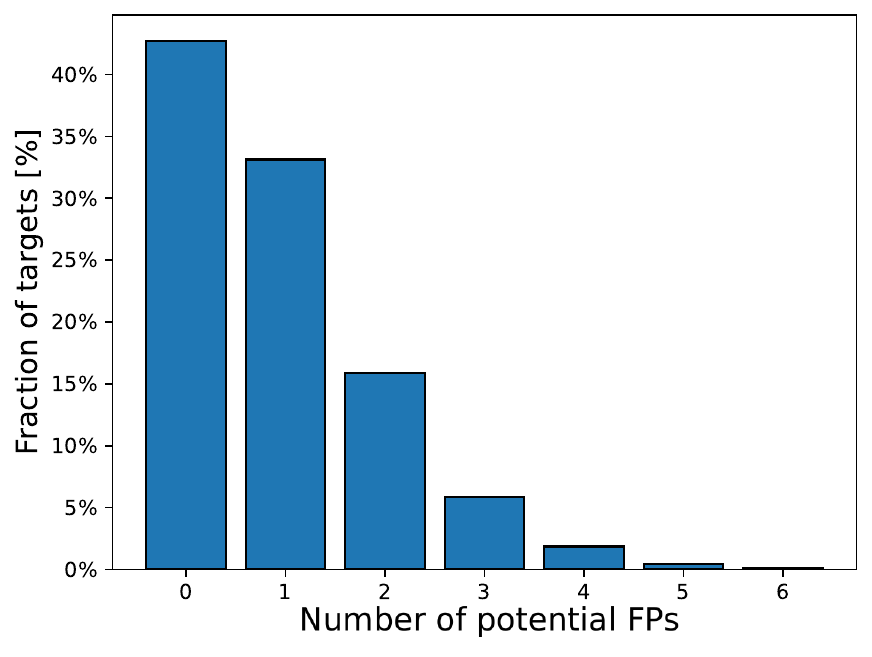}
    \caption{Distribution of the number of contaminants with $\rm \eta_{k}^{nom} >\etamin$ for the targets in the magnitude range $10 \leq \rm P \leq 13$ from the catalog described in Sect. \ref{sec:methods} and observed with 24 cameras.}
    \label{fig:fps}
\end{figure}

We now study the number of unique mask shapes for the nominal, extended and secondary masks. This is important because the introduction of extended and secondary masks could increase the total number of mask shapes such that the allowed, on-board limit of mask shapes is reached. In more detail, the on-board software has a mask library that consists of 8,000 different shapes that are shared among the three types of masks. This library is prepared on-ground and it is programmed to be shared between the 24 N-CAMs. Fig.~\ref{fig:cummul_masks} shows the cumulative number of unique masks for the extended, nominal and secondary masks. The total number of unique mask shapes is 1179. This is way below the 8,000 threshold in the mask library.
\begin{figure}[htbp]
        \centering
        \includegraphics[width=0.95\columnwidth]{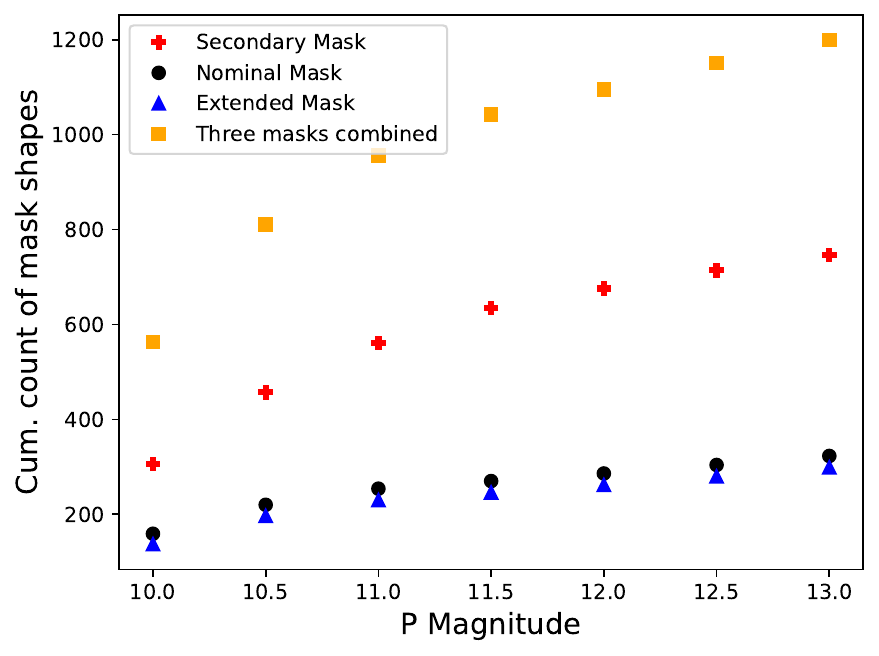}
        \caption{Cumulative count of unique mask shapes as a function of target P magnitude. The plot shows the unique number of each mask, i.e. extended (blue triangles), secondary (red crosses) and nominal (black circles) and also the three types of masks together (orange squares).}
        \label{fig:cummul_masks}
    \label{fig:masks_comparison}
\end{figure}

\subsection{Efficiency of the metrics}
\label{subsec:efficiency_of_the_metrics}

Figures~\ref{fig:comp_cob_shift_variable} and~\ref{fig:comp_dap_flux_variable} show the results for the efficiency of centroid shift and flux measurements using nominal and double-aperture photometry. 
 To produce these plots, we divided the target magnitude range (10 $\leq \rm P \leq$ 13) into 7 magnitude bins of 0.5 magnitudes in size. 
  
For each bin, we randomly select 1000 targets from all available stars within the range $\rm P \pm 0.25$ magnitudes around each central value.  Each data point represents the efficiency calculated for all 1000 targets within that specific magnitude bin. Also, all magnitude values referenced in the efficiency plots and tables correspond to target star magnitudes, including for secondary mask analyses. 
While secondary masks are centered on the most significant contaminant star, the efficiency is evaluated as a function of target magnitude to maintain consistency across all detection methods and to align with the magnitude-binned target selection approach described in Sect.~\ref{sec:Efficiency}.

Both Figures show a light blue and purple regions that overlap each other in the magnitude range $\rm 10.7 \leq P \leq 11.7$. The light blue ones show the magnitude region where Earth-like planets can be detected by \plato using 24 cameras. The light purple regions show the magnitude range where photometry extraction is expected to be performed on-board. The region where they overlap is a rough estimate of where, in principle, \plato could detect Earth-like planets with on-board photometry using 24 cameras. The $\rm P=10.7$ (i.e. V = 11 for a star with $\rm T_{eff} = 6000~K$) threshold can be found in both the \PLATORedBook\ and \cite{ESA_2021}. 
On the other hand, the $\rm P=11.7$ threshold is the result of the calculation presented  in Appendix \ref{sec:Appendix_A}, which confirms the results shown in Fig.~18 of \cite{2019Marchioripaper}.

Figure~\ref{fig:comp_cob_shift_variable} shows the efficiency of the centroid shift measurements using both nominal and double-aperture photometry. The calculations were performed for the cases of 24 and 6 cameras. The nominal centroids are very close in efficiency to the extended ones. The secondary centroid shifts are the next ones in efficiency. Since secondary masks are always centered on the most significant contaminant star of each window and if there is a transit happening on that contaminant star, the resulting centroid shift will be small. This means a less significant signal according to Eq.~(\ref{eqn:eta_secondary_centroid_shift}). We also notice a small decrease of their efficiency with increasing P magnitude. This can be explained showing the variation of the averaged centroid shift uncertainty, $\rm \sigma_{\rm k}^{\rm 1h, N_{T}}$, of each metric with the magnitude. Fig.~\ref{fig:cob-noise} shows this variation and we can see that the uncertainty associated with the extended centroid increases more rapidly than does the  uncertainty associated with the nominal centroid. This hence explains the decrease of the efficiency with the magnitude.

Figure~\ref{fig:comp_dap_flux_variable} shows the efficiency of flux measurements using the double-aperture photometry method. The calculations were performed again for both the cases of 24 and 6 cameras. The secondary flux is the most efficient method for both 24 and 6 cameras. This result is intuitive given that secondary masks are always centered on the contaminant with the highest probability to cause a FP in every window, given it has the highest $\rm SPR_{k}$ value and --~in most cases~-- lowest NSR too.

We observe that efficiency is consistently higher for 24 cameras than for 6 cameras across all detection methods. This occurs because photometric uncertainty scales as $\sigma \propto 1/\sqrt{\rm N_{\rm T}}$ for both centroid and flux measurements, where $\rm N_{\rm T}$ is the number of contributing cameras. Therefore, increasing from 6 to 24 cameras reduces uncertainty by a factor of $\sqrt{24/6} = 2$, correspondingly increasing statistical significance by the same factor. A detailed analysis is provided in Appendix~\ref{sect:appx:24_to_6} using the extended flux as example.

\begin{figure*}[htbp]
    \centering 
    \begin{subfigure}[b]{0.46\textwidth}
        \includegraphics[width=\textwidth]{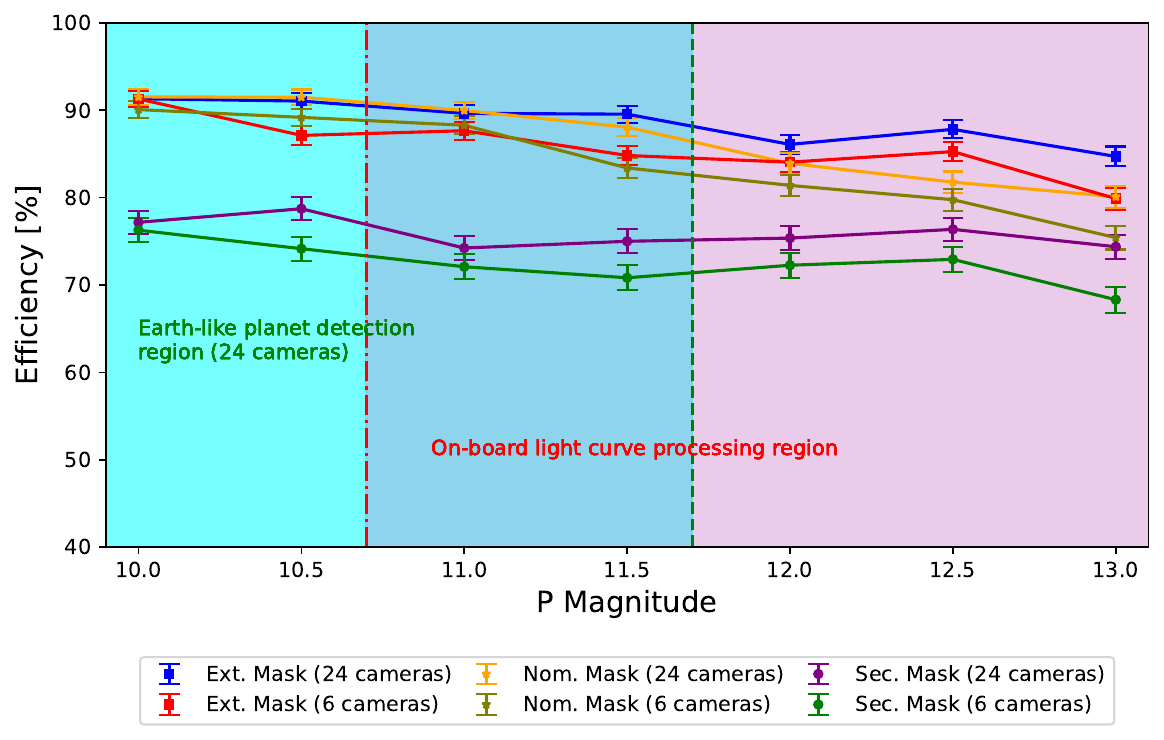}
        \caption{Centroid shift efficiency}        \label{fig:comp_cob_shift_variable}
    \end{subfigure}
    \begin{subfigure}[b]{0.47\textwidth}
        \includegraphics[width=\textwidth]{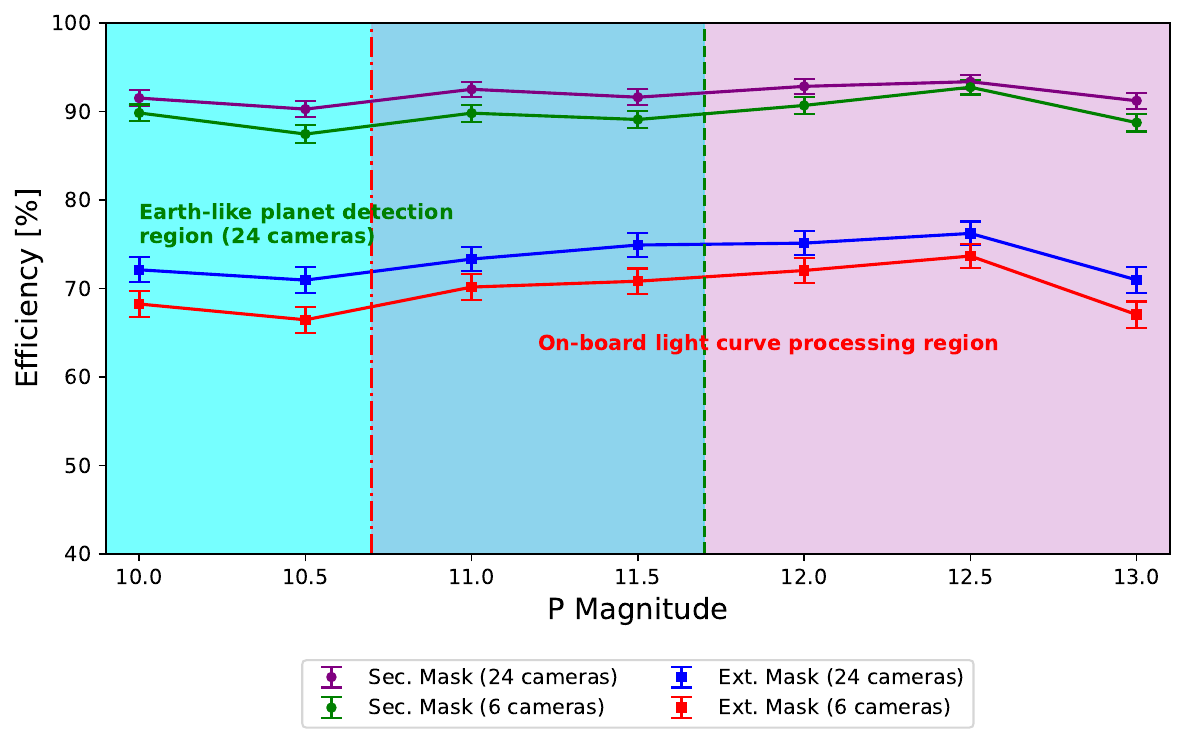}
        \caption{Double-aperture photometry efficiency}
        \label{fig:comp_dap_flux_variable}
    \end{subfigure}
    \caption{Comparison of centroid shifts and flux measurements with double-aperture photometry for detecting FPs. The results are produced for the case of variable transit parameters for the EBs. On the left we see the efficiency for centroid shift measurements using the three types of masks mentioned in this work so far. On the right, the efficiency for both secondary and extended masks. Both figures were obtained using 24 and 6 cameras for each mask. The Earth-like planet detection region is colored in light blue. The vertical red, dot-dashed line corresponds to the P = 10.7 magnitude threshold. The pink colored region is is the region where on-board light curves will be produced on-board for the P5 sample. The threshold of P = 11.7 is the vertical green, dashed line.}
    \label{fig:comp_methods_variable}
\end{figure*}

Table~\ref{tab:efficiency_comparison} shows the magnitude-averaged efficiencies across the entire P5 range of the values in Figures~\ref{fig:comp_cob_shift_variable} and~\ref{fig:comp_dap_flux_variable}. 
\begin{table}[htbp]
\centering
\caption{Detection averaged efficiency}
\label{tab:efficiency_comparison}
\begin{tabular}{lc}
\hline
\textbf{Metric} & \textbf{Efficiency}  \\
\hline
Nominal centroids (NCOB) & $83.7 \pm 1.2\%$ \\
Extended centroids  (ECOB) & $87.1 \pm 1.0\%$  \\
Secondary centroids (SCOB) & $75.4 \pm 1.4\%$ \\
Secondary flux (SFX) & $92.1 \pm 0.8\%$ \\
Extended flux (EFX) & $73.5 \pm 1.4\%$ \\
Extended flux (EFX with $\rm \delta_{k}^{ext} > \delta_{k}^{nom}$)  & $87.2 \pm 1.0\%$ \\
\hline
\end{tabular}
\tablefoot{Detection averaged efficiency across the entire P5 magnitude range.}
\end{table}

The results present in Table~\ref{tab:efficiency_comparison} show that secondary flux is the most efficient metric for detecting  FPs. It is followed by the  centroid shift measurements and finally by the extended flux. Secondary flux is particularly effective when only one contaminant is capable of generating a FP, as secondary masks are specifically designed for single-contaminant scenarios. However, when multiple contaminants can generate FPs, nominal and extended centroids become preferable due to their ability to simultaneously monitor several contaminants and their higher efficiency compared to extended flux, though only where computational resources allow it. Furthermore, flux measurements involving double-aperture photometry are both CPU and telemetry cheaper than centroid shift measurements computed either with a  nominal or secondary or extended masks. This is because flux measurements require only a single scalar value per aperture, while centroid measurements must compute and transmit both x and y coordinates. Taken together, these considerations suggest that for the targets where flux measurements are as efficient as centroid shift measurements, the former should be preferred over the latter. More detailed suggestions about this are given in Sect.~\ref{subsec:overall_strategy_for_selecting_the_best_metrics}.

\begin{figure}[htbp]
        \centering        \includegraphics[width=0.95\columnwidth]{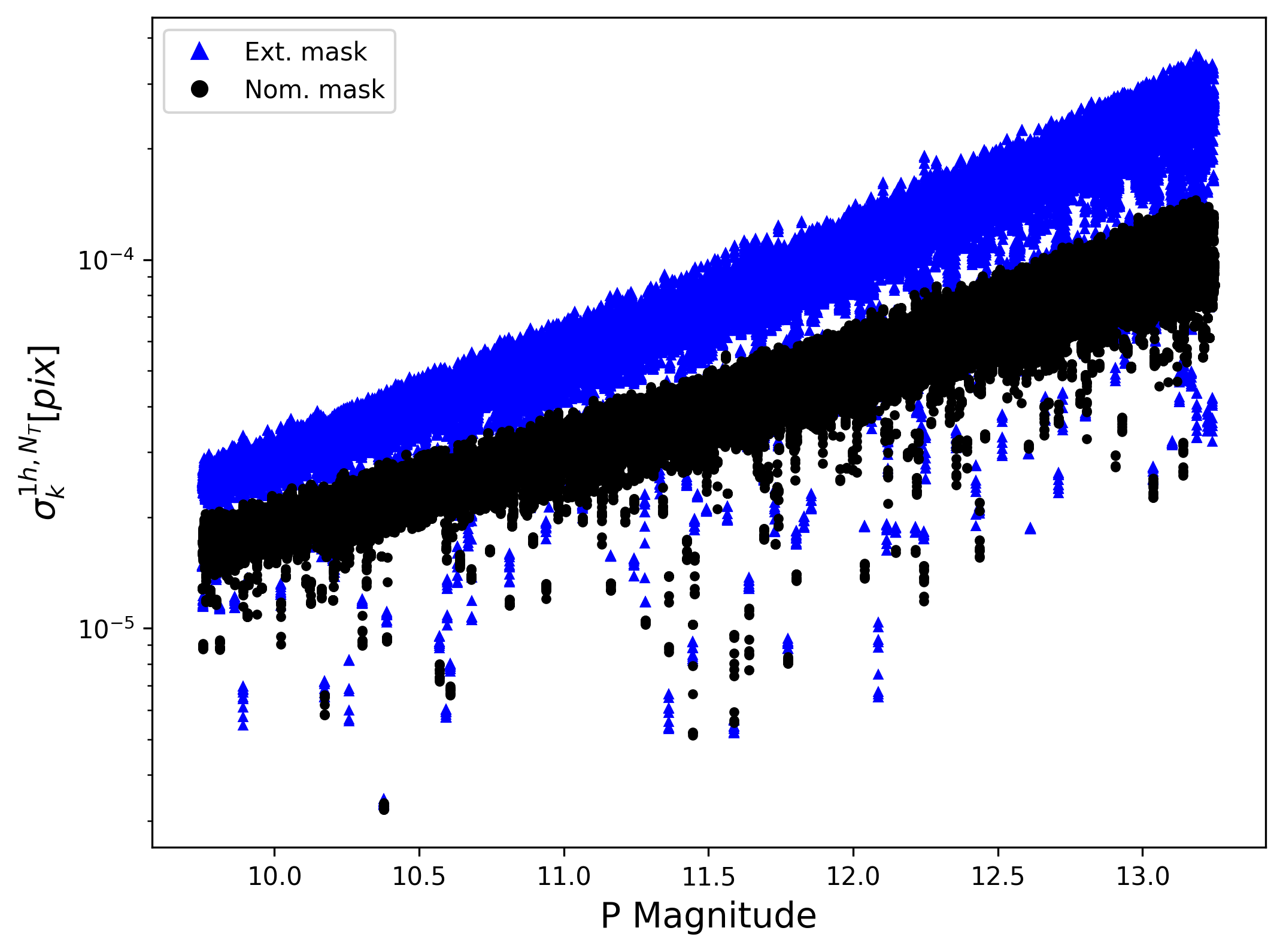}
        \caption{ Distribution of the centroid shift averaged uncertainty (Eq.~(\ref{eqn:centroid-noise})) along target P magnitude for the extended (blue triangles) and nominal (black circles) masks.}
        \label{fig:cob-noise}
\end{figure}

\section{Discussion and conclusions}
\label{sec:Conclusions}
\subsection{Overall efficiency of the different metrics}
\label{subsec:overall_efficiency_of_the_different_metrics}

As discussed in the previous sections, the efficiency of the different metrics depends on the conditions under which they are evaluated (see Sect.~\ref{sec:Efficiency}). These conditions are related to how \plato measurements are obtained. Any change in the conditions will create changes in the efficiency of the metrics. An illustrative example is given by the extended flux measurements: if we replace the condition ($\rm \delta_{k}^{ext} > \delta_{k}^{nom} + 3 \, \sigma_\delta$) by the condition ($\rm \delta_{k}^{ext} > \delta_{k}^{nom}$) we get an efficiency for the extended flux of: 87.2\,\%. As pointed out in Sect.~\ref{sub_subsec:extended_mask},
in the expression for $\rm \sigma_\delta$ given by Eq.~(\ref{eqn:sigma_delta}), we have assumed that the analysis of both the extended and  nominal flux  are done independently. A differential analysis of the extended flux and the nominal flux, as mentioned in the same Section, is expected to substantially reduce $\rm \sigma_\delta$ and make the detection of FPs with the extended flux more efficient. Indeed, in the limit case where $\rm \sigma_\delta$  tends to zero, the condition $\rm \delta_{k}^{ext} > \delta_{k}^{nom} + 3 \, \sigma_\delta$ tends to the condition $\rm \delta_{k}^{ext} > \delta_{k}^{nom}$. Hence, a differential analysis  should increase the efficiency of the extended flux and put the latter somewhere between 73.5\,\% and 87.2\,\%. However, it should be noted that  a differential analysis would require dedicated processing to be implemented on ground, which is not currently planned for the \plato Exoplanet Analysis System (EAS). 
   
\subsection{ Overall strategy for selecting the best metrics }
\label{subsec:overall_strategy_for_selecting_the_best_metrics}

Just after the secondary flux, the extended and nominal centroid measurements are the most efficient method for detecting FPs. However, the maximum number of centroids that can be measured on-board is limited due to the CPU, memory and telemetry constrains. Indeed, we have the following on-board constraints (the numbers below are given for a single camera):
\begin{itemize}
    \item Max. number of nominal centroids per camera: 7,400 ;
     \item Max. number of extended centroids per camera: 7,400 ;
    \item Max. number of extended or secondary flux per camera: 45,400;
    \item Max. number of nominal flux per camera: 104,850 \footnote{This number includes the stars from the guest observer program, that is about 20,000 targets per camera and for a single pointing.}.
\end{itemize}
 
Within these limits,  it is possible to consider an optimal distribution between the centroid measurements and the extended/secondary flux measurements. To establish this share it is required to isolate the cases each metric is able to detect a FP while another cannot. We care about the number of targets per camera now because this is how most CPU resources are organized in the pipeline.\\

We present now the percentage of FPs detected by each metric in combination with the other ones. The following abbreviations were introduced for convenience:  NCOB for nominal centroid shift measurements, ECOB for extended centroid shift measurements, EFX for extended flux measurements, SFX for secondary flux measurements and SCOB for secondary centroid shift measurements. Table~\ref{tab:fp_detection_percentage_matrix_10contaminants} shows the percentage of FPs for the different combinations of extended and nominal mask-based methods. For these comparisons the first 10 most significant contaminants in terms of $\sprk$ in each window were considered. Accordingly, Table~\ref{tab:secondary_methods_comparison} shows the FPs percentage detection among the combinations of secondary mask-based methods. For these comparisons only the most prominent contaminant in terms of $\sprk$ of each window was considered. \footnote{The fractions of FPs detection were computed by considering the proportion of contaminant stars that meet the detection criteria for each metric described in Sect. \ref{sec:Efficiency} in each magnitude bin. These fractions were then weighted based on the actual distribution of stars across magnitude bins, with error estimates derived from the weighted variance assuming a binomial distribution.}:

\begin{table}[htbp]
\centering
\caption{Percentage of FPs detectable by a method but not by another.}
\label{tab:fp_detection_percentage_matrix_10contaminants}
\begin{tabular}{|c|c|c|c|}
\hline
     & \textbf{NCOB} & \textbf{ECOB} & \textbf{EFX} \\
\hline
\textbf{NCOB} & --- & $4.2 \pm 0.6$ & $17.4 \pm 1.2$ \\
\hline
\textbf{ECOB} & $7.6 \pm 0.8$ & --- & $13.9 \pm 1.1$  \\
\hline
\textbf{EFX} & $7.2 \pm 0.8$  & $0.3 \pm 0.2$ & --- \\
\hline
\end{tabular}
\tablefoot{Percentage of FPs detectable by the method in the row but not by the method in the column, using the 10 most significant contaminants per window.}
\end{table}

\begin{table}[htbp]
\centering
\caption{Percentage of FPs detectable by a method but not by another.}
\label{tab:secondary_methods_comparison}
\begin{tabular}{|c|cc|}
\hline
     & \textbf{SFX} & \textbf{SCOB} \\
\hline
\textbf{SFX} & --- & $20.6 \pm 1.3$ \\
\hline
\textbf{SCOB} & $3.9 \pm 0.6$ & --- \\
\hline
\end{tabular}
\tablefoot{Percentage of FPs detectable by the method in the row but not by method in the column, using only the most prominent contaminant per window (secondary masks).}
\end{table}

Table~\ref{tab:fp_detection_percentage_matrix_10contaminants} shows that extended and nominal centroid shifts detect a similar fraction of FPs. Nominal centroid shifts recover $17.4\,\%$ of FPs missed by extended fluxes, while extended centroid shifts recover $13.9\,\%$. Conversely, extended fluxes identify $7.2\,\%$ of FPs missed by nominal centroid shifts but only $0.3\,\%$ of those missed by extended centroid shifts. These results highlight the strong complementarity between extended fluxes and both extended and nominal centroid shifts. 
 Accordingly, the choice between the extended centroids, nominal centroids and extended fluxes have to be made  such as to optimize both the CPU and TM resources and the amount of FPs potentially detectable for each given target. However, extended fluxes could be prioritized since they have a lower cost in CPU and TM; also, nominal centroids could be prioritized over the extended centroids because they can be directly implemented in a window without invoking an additional, extended aperture. 
Table~\ref{tab:secondary_methods_comparison} shows that secondary flux clearly outperforms secondary centroids, detecting $20.6\,\%$ of FPs that secondary centroids cannot identify, while secondary centroids detect only 3.9\,\% of FPs missed by secondary flux. This demonstrates the greater superiority of flux-based over centroid-based detection when focusing on the most problematic contaminant. Together with the multi-contaminant comparison above, these results highlight the complementarity of flux- and centroid-based metrics, and how their usefulness depends on the context of each window and the available CPU and TM resources.

Now we can propose a strategy to select the best metric for each target, since this is how metrics will be assigned on ground (per target). The strategy shall take into account the number of contaminants able to create a FP in each window, $\rm N_{FP}$, as well as the results and computations presented so far. The strategy is summarized as follow
\begin{itemize}
    \item $\rm N_{FP} \geq 2$: for each target, we choose between the extended flux, the nominal and extended centroid  metrics based on which metric detects the larger number of FPs. If for a target the extended flux is able to detect the same number of FPs than the  centroids, we shall choose the former because of its lower cost, otherwise we select the nominal centroid, and if needed the extended centroid ;
    
    \item $\rm N_{FP} = 1$ : we choose between the secondary flux and the nominal centroid. When for a given target 
    the FP is detectable by the secondary flux (i.e. if the condition of Eq.~(\ref{eqn:sec_condition}) is verified) we select this metric  because of its lower cost, otherwise  we select  the centroid metric ;
    
    \item $\rm N_{FP} = 0$: our proposal is to choose --~ if the resources allow it~-- an extended flux by default.  This is because  we might have missed some contaminant stars  or we may have wrong information about them. By choosing by default an extended flux we hence further maximize our chance to detect a FP.

 \end{itemize}

Table~\ref{tab:scenarios} summarizes the previous list.
\begin{table}[htbp]
    \centering
    \caption{Proposed strategy to assign metrics}
    \begin{tabular}{|c|c|}
    \hline
    $\rm N_{FP}$ per window &  Metric\\
    \hline
    0 & EFX \\
    \hline
    1 & SFX or NCOB\\
    \hline
    2 or more & EFX or NCOB or ECOB \\
    \hline
    \end{tabular}
    \tablefoot{Decision scheme about different scenarios involving the optimal metric to choose depending on the number of contaminants able to create a FP in each window. }
    \label{tab:scenarios}
\end{table}

\subsection{An example of metric assignment}
\label{subsec:an_example_of_metric_assignment}

To assign a metric to a target we have to take into account for limitations imposed by the on-board software or the TM budget. At the same time, it has to take the scientific requirements into account as well. For instance  the number of P5 targets for which photometry can be extracted on-board is 80, 000 for a single N-CAM. At the same time, centroid computations are limited to 14,800  targets per N-CAM while the number of extended/secondary flux to 45,400 targets again for a single N-CAM. 
 
Table~\ref{tab:assigment} shows our summarized proposed metric assignment scenario. To build it, we started like follows: From Fig.~\ref{fig:fps} we see that the number of P5 targets with  $\rm N_{FP} = 1$ is $\sim 35 \,$\%, which corresponds to about 28,000 targets for a single N-CAM. Given that the efficiency of the secondary mask is 92\,\%, we consider a secondary mask for the 92\,\% of the 28,000 targets for a single N-CAM, these are about 25,760 targets. These are the targets that shall have a secondary flux computed on-board whereas for the remaining  $\sim 2,240$ stars, a nominal centroid measurement shall be considered. These values correspond to the $\rm N_{FP}=1$ row in Table~\ref{tab:assigment}. Furthermore, Fig.~\ref{fig:fps} shows that among the P5 sample, $\sim$ 22 \% are expected to have $\rm N_{FP} \geq 2$ ; this represents a total of $\sim 17,600$~targets for a single N-CAM. These targets in principle can  be monitored either with extended flux measurements, nominal or extended centroid shift measurements. We can use Table~\ref{tab:fp_detection_percentage_matrix_10contaminants} to find a possible share between metrics. According to it, extended centroid shifts detect 7.6\,\% of the FPs missed by nominal centroids, while nominal centroids detect 17.4\,\% of the FPs not detectable by extended fluxes. Therefore, we can start by assigning an extend centroid to 1,337 targets (7.6\,\% of 17,600) for which $\rm N_{FP} \geq 2$. Then, we can assign a nominal centroid to 2,829 targets ((17,600 - 1337) $\times$ 0.174). Lastly, the remaining targets can get an extended flux, that is 13,400 targets (17,600 - 1,337 - 2,829). These values correspond to the $\rm N_{FP}>2$ row of Table~\ref{tab:assigment}. We recall now that the total centroid budget for a single N-CAM is up to 14,800 (7,400 nominal + 7,400 extended). Finally, we can say that the maximum number of extended/secondary flux measurements (45,400) together with the maximum number of centroid measurements (14,800) fully suffices for all the targets for which  $\rm N_{FP} > 0$. The remaining number of extended flux measurements (6,396 of the 45,400) can be assigned to the remaining P5 targets for which  $\rm N_{FP} = 0$ (the $\sim 40$\,\% of the P5 sample) ; these targets can for instance be selected by putting more priority to the brightest ones. This gives the $\rm N_{FP}=0$ row in the Table~\ref{tab:assigment}. To conclude, this metrics share results 
in an optimal use of the resources while fulfilling the technical constraints: we have 25,760 secondary flux,  19,830 extended flux and 6,433 centroid measurements leading to a total of 52,023 targets, staying within the 80,000 targets for a single N-CAM budget, as summarized in the last row of Table~\ref{tab:assigment}.

\begin{table*}[htbp]
    \centering
    \caption{Possible scenario for the metrics}
    \begin{tabular}{|c|c|c|c|c|c|}
    \hline
    $\rm N_{FP}$  & Num. of stars & Secondary flux & Extended flux & Nominal centroid & Extended centroid\\
    \hline
    0 &  34,400 [43 \%] &   -  & 6,396 [8.0\,\%] &  -  & - \\
    \hline
    1 & 28,000 [35 \%] &  25,760 [32.2\%] &  -  &  2,240 [2.8\%]  &  -\\
    \hline
   > 2 & 17,600 [22 \%]  & - &  13,434 [16.7\%] &  2,829 [3.5\%] & 1,337 [1.7\%]\\
       \hline
       \hline
Total: & 80,000  [100 \%] &   25,760 [32.2\%]  &  19,830 [24.8 \%]  & 5,069 [6.4\%]   & 1,337 [1.7\%]\\ \hline
    \end{tabular}
    \tablefoot{Possible scenario for metrics assignments. The quantities inside brackets refer to the percentages in terms of the maximum 80,000 targets for a single N-CAM. }
    \label{tab:assigment}
\end{table*}

\subsection{Concluding remarks}
\label{subsec:concluding_remarks}

This work evaluates the efficiency of double-aperture photometry for detecting FPs. With this new method we produced flux measurements that were compared  with the known centroid shift method in terms of FPs detection. The secondary flux was shown to be the most efficient metric overall; it is followed in efficiency by extended centroid shifts, then by the nominal centroids and then by the secondary centroid shifts. The extended flux is the last metric in terms of overall efficiency. However, we showed that for a large fraction of the \plato P5 targets, flux measurements using the double-aperture photometry are an alternative that is more competitive in terms of CPU and memory resources than the centroid shift method.  We then established a strategy to optimally detect FPs with the double-aperture photometry and centroid methods taking into account the technical limitations of the \plato software and hardware. We finally conclude that by using the double-aperture photometry approach alongside centroid shift measurements implemented on-board,  a large fraction of FPs are expected to be discarded for the \plato P5 sample. Our approach can be used further as a guideline for the design of future space missions aiming at detecting exoplanets with the transit technique. 

A future work involving a differential analysis for extended and nominal masks is the next step. Furthermore, we think another natural step is to analyze the dependence of our results on the following aspects: different FoV, camera pointings and color-dependency of the targets and contaminants. We can refer however to the work by \cite{gutierrezcanales2025}\footnote{The thesis is available here: \url{https://theses.hal.science/tel-05165095}.} for a detailed analysis of the metrics efficiency under different values of PSF temperature, diffusion kernel and other configurations. Finally, a possible improvement for extended masks could consist in computing them such as to maximise the significance (for both extended flux and extended centroid shift measurements) of all the contaminant stars that can potentially generate a background transit. The exact procedure to obtain these new, optimal extended masks needs, however, to be studied in the future.

\begin{acknowledgements} This work presents results from the European Space Agency (ESA) space mission PLATO. The PLATO payload, the PLATO Ground Segment and PLATO data processing are joint developments of ESA and the PLATO Mission Consortium (PMC). Funding for the PMC is provided at national levels, in particular by countries participating in the PLATO Multilateral Agreement (Austria, Belgium, Czech Republic, Denmark, France, Germany, Italy, Netherlands, Portugal, Spain, Sweden, Switzerland, Norway, and United Kingdom) and institutions from Brazil. Members of the PLATO Consortium can be found at \url{https://platomission.com/}. The ESA PLATO mission website is \url{https://www.cosmos.esa.int/plato}. We thank the teams working for PLATO for all their work.
This work has benefited from financial support by  Centre National d'Etudes Spatiales (CNES)
in the framework of its  contribution to the PLATO mission. 
FGC thanks the CNES and the Max Planck Institut für Sonnensystemforschung for its financial support during his PhD program. This work was supported in part by the German space agency (Deutsches
Zentrum für Luft- und Raumfahrt) under PLATO grant 50OP1902. This research has made use of astropy \citep[see
][]{astropy_2022}. FGC thanks Jesper Schou for his valuable comments and help, specially regarding the error bars in the efficiency plots. FGC also thanks Matthias Ammler-von Eiff and René Heller for their valuable comments and Ilyas Kuhlemann for his valuable help in the python code. 
\end{acknowledgements}

\bibliographystyle{aa}

\bibliography{biblio}

\appendix

\section{Planets detectability}
\label{sec:Appendix_A}
Although the main interest behind \plato mission is to discover Earth-like planets, the mission is capable of detecting different types of planets as well. We are interested now in studying the planet detectability for three different types of planets for \plato:  Jovian, Super-Earths and Earth-like planets.  Fig.~\ref{fig:comp_etas} shows our results for each one of these planets. The results were produced taking into account $\rm n_{tr} = 3$.
\begin{figure}[htbp]
    \centering
    \includegraphics[width=0.85\columnwidth]{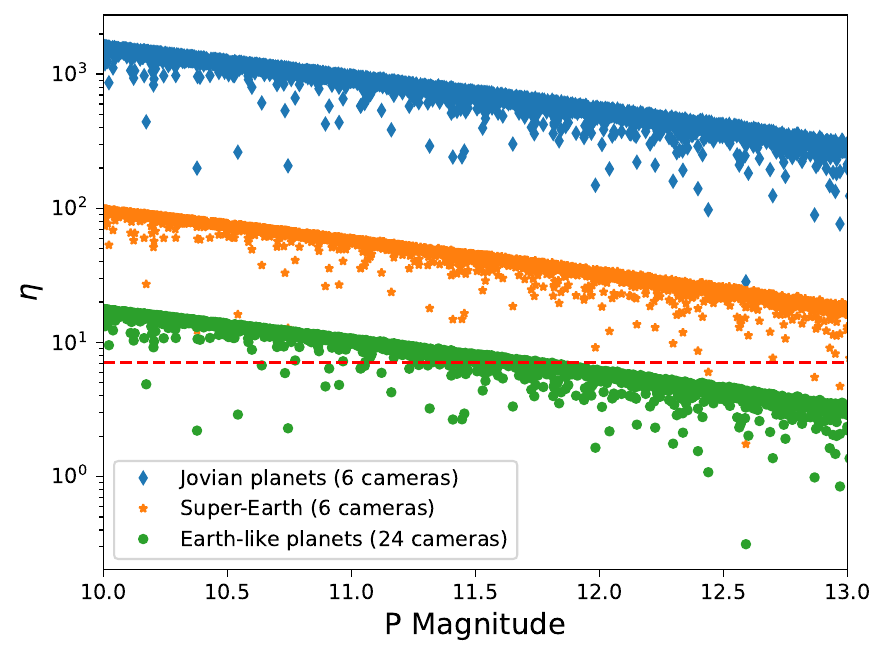} % PDF version also available
    \caption{Statistical significance of true planets (Eq.(\ref{eqn:significance_true_planet})), for three different types. The results are computed for 10000 targets in total. The red, dashed line refers to the value of the detection threshold $\etamin$. The blue diamonds refer to the significance values related to Jovian planets using 6 cameras ($\deltap = 10100$ ppm and $\rm t_{d} = 29.6$ hr), the orange stars refer to Super-Earths using 6 cameras ($\deltap = 522$ ppm and $\rm t_{d} = 42$ hr), and the green circles refer to Earth-like planets using 24 cameras ($\deltap = 84$ ppm and $\rm t_{d} = 13$ hr). The chosen $\deltap$ and $\rm t_{d}$ values for the Jovian and Earth-like planets come from Fig.~18 from~\cite{2019Marchioripaper}. The chosen $\deltap$ and $\rm t_{d}$ values for the Super-Earths were estimated interpolating between the Earth and Neptune values reported in Table 1 from~\cite{Borucki1996}. } 
\label{fig:comp_etas}
\end{figure}

Our analysis show that the significance of Earth-like planets around Sun-like stars has a magnitude limit at $\rm P \sim 11.7$, using $\rm N_{T} = 24$ and $\rm n_{tr} = 3$. This is consistent with the results provided by Fig. 18 from \cite{2019Marchioripaper}. Another consistent result from Fig.~\ref{fig:comp_etas} is that PLATO is able to detect Jovian planets even if only using $\rm N_{\rm T} = 6$. In the case of Super-Earths, there was no previous analysis by \cite{2019Marchioripaper}, but our results show as well that only using $\rm N_{T} = 6$ could be enough for PLATO to detect Super-Earths since they appear to be well above the detectability threshold $\etamin$ when using 6 cameras. 

\section{Nominal mask computation}
\label{sec:Apendix_B}

The NSR expression useful to obtain the nominal mask is Eq. (36) from \cite{2019Marchioripaper} 
\begin{equation}
    \rm NSR_{n} =  \frac{\sqrt{ {\rm I^{T}_{n} } + \sum\limits_{\rm k=1}^{\rm N_{C}} { \rm I^{k}_{n}} + {\rm B} \, \Delta \rm t_{\rm exp} + \sigma^{2}_{\rm D} + \sigma^{2}_{\rm Q} }  }{I^{T}_{n}} \; .
    \label{eqn:NSR}
\end{equation}
 The terms used in Eq.~(\ref{eqn:NSR}) are defined in Table \ref{tab:symbols} since they are the same than the ones from Eq.~(\ref{eqn:nsr_1h_nt_cameras}).
The subscript ``$\rm n$'' indicates that we compute the NSR for every window pixel. Once we computed it for 36 pixels of the window, we sort them in increasing order. We use the subscript ``$\rm m$'' for denoting the index of the sorted window pixels. These pixels are used to compute what \cite{2019Marchioripaper} called the ``aggregate NSR", which is written as
\begin{equation}
    \rm NSR_{agg} (m) = \frac{\sqrt{ \sum\limits_{\rm n=1}^{\rm m} \left( {\rm I^{T}_{n} } + \sum\limits_{\rm k=1}^{\rm N_{C}} { \rm I^{k}_{n}} + {\rm B} \, \Delta \rm t_{\rm exp} + \sigma^{2}_{\rm D} + \sigma^{2}_{\rm Q}       \right)  }}{\sum\limits_{n=1}^{m} \rm I_{n}^{T}} \; .
    \label{eqn:aggnsr}
\end{equation}
We finally obtain the nominal mask, $\omega_{\rm n}$, by looking for the collection of pixels that minimizes Eq.~(\ref{eqn:aggnsr}). This is shown in Figs. 13 and 14c of \citet{2019Marchioripaper} and in Figs. \ref{fig:nsr_agg_1h} and \ref{fig:nominal_mask} of this work. This refers to a star with $\rm P = 11.12$.

\begin{figure}[htbp]
    \centering
    \begin{subfigure}[b]{0.75\columnwidth}
        \hspace{-1cm} 
        \includegraphics[scale=0.5]{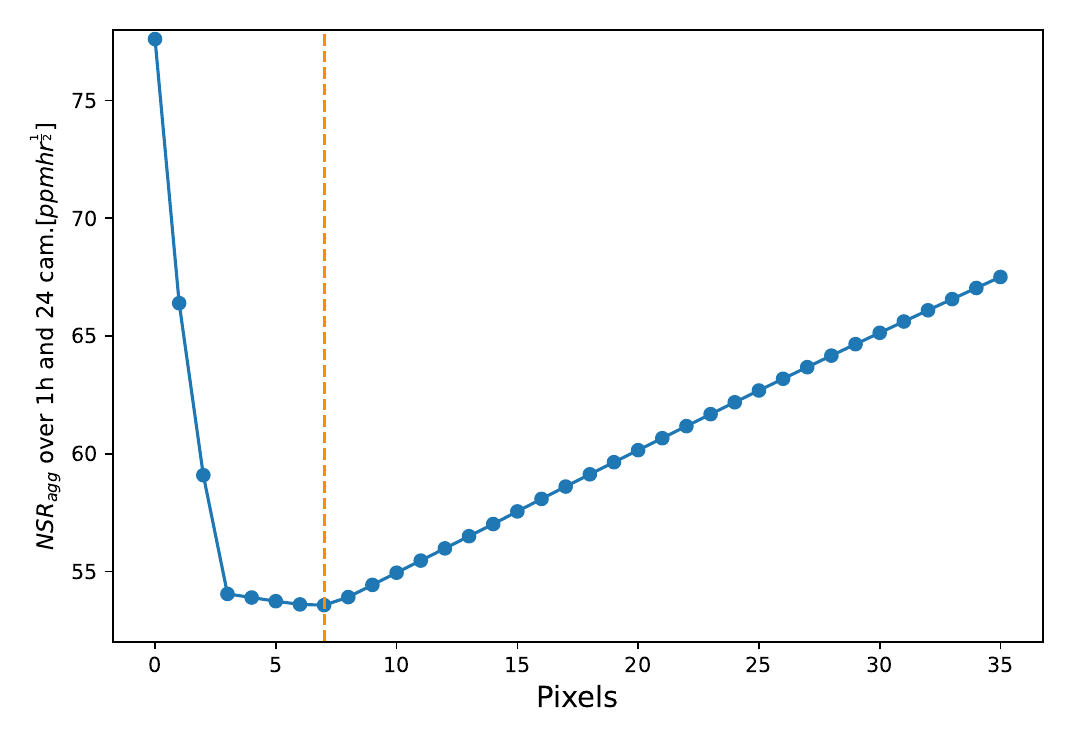}
        \caption{Plot of Eq.(\ref{eqn:aggnsr}). The minimum of the curve indicates the number of pixels (size) of the nominal mask.}
        \label{fig:nsr_agg_1h}
    \end{subfigure}
    
    \begin{subfigure}[b]{0.75\columnwidth}
        \includegraphics[width=\textwidth]{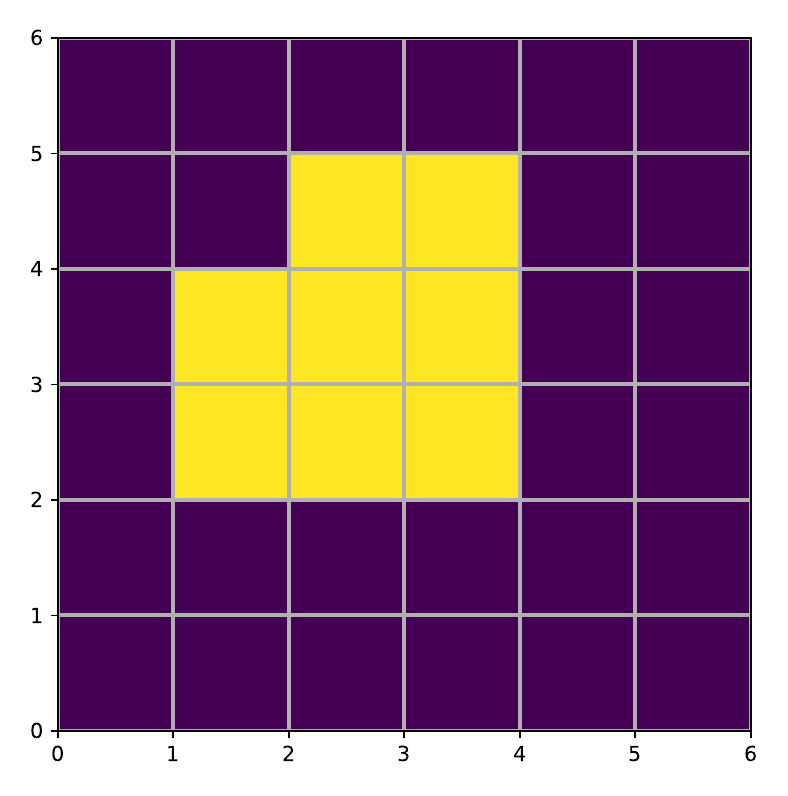}
        \caption{Nominal mask. The purple color indicates a value of 0 and the yellow color indicates a value of 1.}
        \label{fig:nominal_mask}
    \end{subfigure}
    \caption{Top: Example of NSR plot and Bottom: the consequent shape of the nominal mask. For this case, the size of the nominal mask is eight pixels.}
    \label{fig:nominal_mask_procedure}
\end{figure}
The way for building a secondary mask is completely analogous to what have just been described. The only difference is that we substitute $\rm I_{\rm n}^{\rm T}$ by $\rm I_{\rm n}^{\rm kmax}$ since for the secondary mask the star with the highest SPR value in the window is regarded as the target. Furthermore, the way of building the extended mask is simply by surrounding the nominal mask by a single pix. Fig.~\ref{fig:mask_size} shows the average size in pixels of the three types of masks as a function of the P target magnitude.
\begin{figure}[htbp]
    \centering
    \includegraphics[width=0.85\columnwidth]{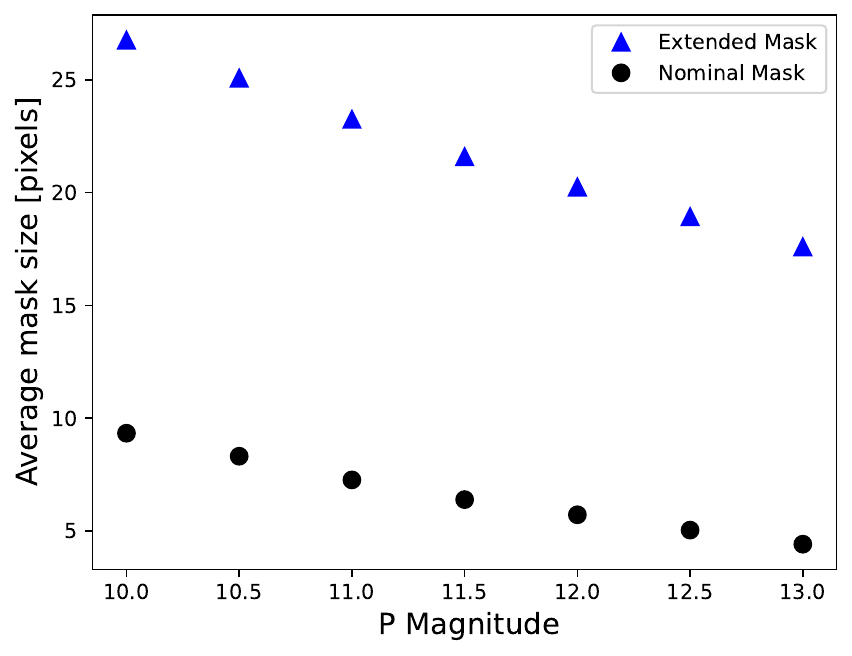}
    \caption{Average mask size in pixels for the nominal (black circles) and extended (blue triangles) masks as a function of the target P magnitude.  }
    \label{fig:mask_size}
\end{figure}
    
\section{Change in efficiency from 24 to 6 cameras}
\label{sect:appx:24_to_6}
To investigate the change in efficiency of both double-aperture photometry and centroid shifts when shifting from 24 to 6 cameras, we use the extended mask as an example. The analysis for other versions of both methods is analogous. We begin with the following expression involving $\rm N_{FP}$ and $\rm N_{FP}^{ext}$ for both 24 and 6 cameras:
\begin{equation}
    \rm \frac{N_{FP}^{ext, 6 cameras}}{N_{FP}^{6 cameras}} = \frac{N_{FP}^{ext, 24 cameras}}{N_{FP}^{24 cameras}} \left[ \frac{(1 - \alpha)}{(1 - \beta)}\right] \; ,
    \label{eqn:24_6_cameras_expression}
\end{equation}
where $\rm \alpha$ and $\rm \beta$ are terms useful for investigating the efficiency change with different camera numbers. This change results from the loss of a certain number of stars when switching the camera number. We denote this loss as $\rm Loss^{ext}$, defined as:
\begin{equation}
    \rm Loss^{ext} = \alpha N_{FP}^{ext, 24 cameras} \; .
    \label{eqn:Loss_ext}
\end{equation}
We observe that switching from 6 to 24 cameras can double the signal. The efficiency change can be attributed to the distribution change of $\rm \eta_{\rm k}^{\rm ext}$ (denoted as $\rm L_{1}$), the distribution change of $\rm \eta_{\rm k}^{nom}$ (denoted as $\rm L_{2}$), or a combination of both ($\rm L_{1} + L_{2}$). Additionally, the efficiency change might depend on the denominator, the distribution change of $\rm \eta_{\rm k}^{\rm nom}$ (denoted as $\rm L_{2}$). This is expressed by
\begin{equation}
\begin{aligned}
    \rm L_{1} &= \rm (\etamin < \etaext < 2\etamin) \land (\etanom < 2\eta_{min}) \; ,\\
    \rm L_{2} &= \rm (\eta_{min} < \etanom < 2\etamin) \land (\etaext < 2\etamin) \; , \\
    \rm L_{3} &= \rm L_{1} + L_{2} \; ,\\
    \rm L_{4} &= \rm (\eta_{min} < \etanom < 2 \eta_{min}) \; .
\end{aligned}
\end{equation}
Then equation (\ref{eqn:Loss_ext}) becomes
\begin{equation}
\begin{aligned}
    \rm \text{Loss}^{ext} &= \rm \alpha N_{FP}^{ext, 24 cameras} = \rm \sum\limits_{\text{targets}} \sum\limits_{k = 1}^{N_{C}} (\delta_{k}^{ext} > \delta_{k}^{nom}) \times \\
    &\quad \Bigl[ \rm L_{1} \text{ or } L_{2} \text{ or } L_{3} \text{ and } L_{4} \Bigr] \; .
    \label{eqn:loss_ext_expaned}
\end{aligned}
\end{equation}
We then get the values for $\rm L_{1}$, $\rm L_{2}$, $\rm L_{3}$ and $\rm L_{4}$. From equation \ref{eqn:loss_ext_expaned} we derive:
\begin{equation}
    \rm \alpha = \frac{L_{1} + L_{2} + L_{3}}{N_{FP}^{ext, 24 cameras}} \; ,
    \label{eqn:alpha}
\end{equation}
and
\begin{equation}
    \rm \beta = \frac{L_{4}}{N_{FP}^{ext, 24 cameras}} \; .
    \label{eqn:beta}
\end{equation}
Our analysis shows that $\rm (1 - \alpha)$ is always greater than $\rm (1 - \beta)$

\section{Reference table}
With the aim to make our results as reproducible as possible, the interested reader can have Table \ref{tab:important_metrics} as reference. Table \ref{tab:important_metrics} shows the most relevant metrics for a given target in our stellar sample and the 10 most significant contaminants around it.

\begin{table*}[htbp]
\centering
\caption{Comparison of Important Metrics}
\label{tab:important_metrics}
\begin{tabularx}{\textwidth}{|S[table-format=19.0]|p{1cm}|p{1cm}|p{1cm}|S[table-format=1.3]|S[table-format=1.3]|S[table-format=1.3]|S[table-format=1.3]|S[table-format=1.3]|S[table-format=1.3]|S[table-format=0.2]|}
\toprule
\multicolumn{11}{c}{\textbf{Target Gaia ID DR3: 5552642365761527040}} \\
\midrule
\multicolumn{11}{c}{\textbf{R.A. and Decl. [deg]: 101.84629226792525 and -46.173084507722656}} \\
\midrule
\multicolumn{11}{c}{\textbf{Number of contaminants: 31}} \\
\midrule
\multicolumn{11}{c}{\textbf{Magnitude: 12.08}} \\
\midrule
\ Contams & \ P mag. & {$\Delta x_{\rm CCD}$} & {$\Delta y_{\rm CCD}$} & {$\rm SPR_{k}^{nom}$[ppm]} & {$\eta_k^{\rm nom}$} & {$\eta_k^{\rm ext}$} & {$\eta_{\rm kmax}^{\rm sec}$} & {$\eta_k^{\rm nom, \Delta C}$} & {$\eta_k^{\rm ext, \Delta C}$} & {$\eta_{\rm kmax}^{\rm sec, \Delta C}$} \\
\midrule
5552642370059980800 & 17.624 & -0.057 & -0.375 & 5961.853 & 17.790 & 38.892 & 21.066 & 10.439 & 5.938 & 2.237 \\
\midrule
5552642404418781184 & 16.545 & 1.091 & 0.522 & 5780.367 & 16.965 & 14.636 & & 13.178 & 50.197 \\
\midrule
5552643121676967936 & 19.321 & 0.402 & 1.791 & 166.539 & 0.516 & 2.377 & & 0.448 & 3.447 \\
\midrule
5552643121676965888 & 20.473 & -0.665 & 1.368 & 107.353 & 0.332 & 1.016 & & 0.360 & 1.473 \\
\midrule
5552642365762705408 & 18.646 & -1.264 & -2.374 & 99.234 & 0.282 & 1.210 & & 0.664 & 10.446 \\
\midrule
5552643125970941952 & 20.001 & 0.158 & 1.834 & 91.024 & 0.096 & 0.657 & & 0.241 & 1.703 \\
\midrule
5552642400122460160 & 18.346 & 2.711 & 0.991 & 31.466 & 0.087 & 0.827 & & 0.061 & 1.066 \\
\midrule
5552643121676971008 & 19.007 & 0.070 & 2.646 & 28.131 & 0.020 & 0.166 & & 0.076 & 1.248 \\
\midrule
5552643125973283840 & 16.977 & -2.454 & 0.309 & 7.096 & 0.018 & 5.488 & & 0.020 & 10.212 \\
\midrule
5552642404418783232 & 16.577 & 4.245 & 0.875 & 6.435 & 0.003 & 1.983 & & 0.016 & 0.266 \\
\midrule
\bottomrule
\end{tabularx}
\end{table*}

\end{document}